\documentclass[10pt,journal,compsoc]{IEEEtran}

\usepackage{cite}
\usepackage{amsmath,amssymb,amsfonts}
\usepackage{algorithmic}
\usepackage{graphicx}
\usepackage{textcomp}
\usepackage{xcolor}
\def\BibTeX{{\rm B\kern-.05em{\sc i\kern-.025em b}\kern-.08em
    T\kern-.1667em\lower.7ex\hbox{E}\kern-.125emX}}
\usepackage{array, colortbl}
\usepackage{ulem, comment}
\usepackage{lscape}
\usepackage{longtable}
\usepackage{adjustbox}
\usepackage{slashbox}
\usepackage{csquotes}
\renewcommand{\mkbegdispquote}[2]{\itshape}
\usepackage{soul,color}
\usepackage{tabularray}
\usepackage[justification=centering]{subfig}
\usepackage[many]{tcolorbox}
\usepackage{balance}

\newcommand{\etal}{\textit{et al.}}
\newtcolorbox{boxblock}[2][]{
top=0.15in,left=4pt,right=4pt,bottom=4pt,
fonttitle=\bfseries,
colbacktitle=gray,
colback=gray!5,
colframe=gray!40!black,
enhanced,
attach boxed title to top left={xshift=1.5em,yshift=-\tcboxedtitleheight/2},
boxed title style={size=small},
drop shadow={black!50!white},
title=#2,#1}

\begin{document}

\title{The Different
Faces of AI Ethics Across the World: A
Principle-Implementation Gap Analysis
}

\author{Lionel~Nganyewou~Tidjon and Foutse~Khomh\IEEEmembership{, Senior Member,~IEEE}
\IEEEcompsocitemizethanks{\IEEEcompsocthanksitem The authors are with Polytechnique Montréal, Montréal, QC H3C 3A7, Canada. 
E-mail: \{lionel.tidjon, foutse.khomh\}@polymtl.ca
}
}

\IEEEtitleabstractindextext{

\begin{abstract}
Artificial Intelligence (AI) is transforming our daily life with several applications in healthcare, space exploration, banking and finance. These rapid progresses in AI have brought increasing attention to the potential impacts of AI technologies on society, with ethically questionable consequences. In recent years, several ethical principles have been released by governments, national and international organisations. These principles outline high-level precepts to guide the ethical development, deployment, and governance of AI. However, the abstract nature, diversity, and context-dependency of these principles make them difficult to implement and operationalize, resulting in gaps between principles and their execution. Most recent work analysed and summarized existing AI principles and guidelines but they did not provide findings on principle-implementation gaps and how to mitigate them. These findings are particularly important to ensure that AI implementations are aligned with ethical principles and values. In this paper, we provide a contextual and global evaluation of current ethical AI principles for all continents, with the aim to identify potential principle characteristics tailored to specific countries or applicable across countries. Next, we analyze the current level of AI readiness and current implementations of ethical AI principles in different countries, to identify gaps in the implementation of AI principles and their causes. 
Finally, we propose recommendations 
to mitigate the principle-implementation gaps. 
\end{abstract}

}
\maketitle
\IEEEpeerreviewmaketitle

\section{Introduction}

AI achieved successful results in several domains such as healthcare~\cite{kourou2015machine}, automotive~\cite{kuutti2020survey}, and aviation~\cite{8903554}. The powerful and profound impact of AI on society led to several debates about the principles and values that must guide its implementation and use. In recent years, several ethical AI principles have been released by governments, national, and international organisations. Some of these principles are generic while others are tailored to a specific environment or context. Most generic principles have been designed by international organisations, with the aim to provide abstract and inclusive principles, that will lead the implementation and use of AI in member countries or companies. Contextual principles have been designed by national organizations and governments to take into account the specific AI factors in a given country. For example, no robotics principle is provided for a country with no robotics industry.  
As a consequence, countries around the world are having different principles tailored to their context in addition to generic principles from the international organizations to which they are affiliated. 
This diversity and context-dependency of ethical 
AI principles make them difficult to implement and operationalize; resulting in gaps between the stated principles and their execution. 

Recently, researchers 
have analysed and summarized AI principles proposed on some continents~\cite{floridi2018ai4people, jobin2019global, hagendorff2020ethics, fjeld2020principled}. However, to the best of our knowledge, none of these studies examined the gaps between the principles and their implementation. 
This paper aims to fill this gap in the literature by analyzing ethical AI principles around the world and the possible gaps between the principles and their implementations. We also investigate the root causes of principle-implementation gaps and propose mitigation strategies. 
The analysis of ethical AI principles around the world is important for understanding the characteristics of 
ethical principles tailored to specific countries, and identifying similarities between principles from different countries, as well as similarities between global principles from different continents and those from international organizations such as the Organization for Economic Cooperation and Development (OECD), the Global Partnership on AI (GPAI), and the Group of Twenty (G20). 

In this paper, we provide a contextual and holistic evaluation of current ethical AI principles and their implementations by analyzing a set of 
100 documents and websites containing 100 AI principles from 29 countries in the 6 continents and 121 
implementations of these principles from 14 countries in the 6 continents (e.g., guidelines, standards, tools). The documents are selected based on their reliability, recency, and diversity. 
Results show 
that the most cited principle 
in the countries is Transparency. The principles that cover most countries
(i.e., global principles) are Privacy, Transparency, Fairness, Security, Safety, Responsibility, Accountability, Explainability, Well-being, Human Rights, and Inclusiveness. Results from the implementation analysis show that United States and United Kingdom are more favorable for investment, business, and research in AI with a good growth stability between 2020 and 2021. They also show the fast growing countries in terms of AI readiness and the current implementation tools (e.g., laws, standards, checklists, training, softwares). The principle-implementation gaps include lack of effective implementation tools and lack of practical training courses on AI. 
Recommendations for gap mitigation include inclusiveness and diversity of team roles, education and awareness on ethical values, and use of EthicsOps. 
EthicsOps is the continuous planning, execution, and monitoring of ethics principles in a project lifecycle (including AI project). It ensures that team roles are diverse (i.e., no race and gender discrimination), business requirements are pro-ethically aligned (i.e., meet ethics laws and standards, ethics governance guidelines), ethical controls are enforced (e.g., checklists), and that ethics by design practices are applied~\cite{eu-ethos} during the project lifecycle.
Governments, national, and international organisations can leverage these results to guide their AI initiatives and conduct ethical AI implementations adapted to their context.

The rest of this paper is structured as follows. Section~\ref{sota} reviews the relevant literature. Section~\ref{methodology} describes the methodology of our study. In Section~\ref{evaluation}, we evaluate recent ethical AI principles tailored to specific countries and those that are more generic. Section~\ref{analysis} presents an analysis of current implementations of AI principles. In Section~\ref{mitigation}, we provide recommendations to mitigate the gaps between AI principles and their implementation.
Section~\ref{threat2valid} discusses threats that could affect the validity of the reported results.
Section~\ref{conclusion} concludes the paper and outlines some perspectives.

\section{Related work}\label{sota}

Floridi~\etal{}\cite{floridi2018ai4people} proposed five ethical principles to lead the development and adoption of AI for the benefit of society and presented 20 concrete recommendations to assess, develop, incentivise, and support good AI. However, the proposed recommendations are high-level and mainly focus on AI ethics governance. This work does not evaluate the application of AI ethics in practice (e.g., country AI readiness analysis, principle-implementation gap analysis). 

Jobin~\etal{}\cite{jobin2019global} performed an in-depth analysis of AI guidelines and observed that they are structured around five global principles (transparency, justice and fairness, non-maleficence, responsibility, and privacy). They also identified two important actions to be undertaken by 
the global community: (1) translating principles into practice and (2) seeking harmonization between AI ethics codes (soft law) and legislation (hard law). However, they have not analyzed ethical AI implementations. 

Fjeld~\etal{}\cite{fjeld2020principled} proposed a visualization of AI principles classified by themes (e.g., International Human Rights, Privacy) and sectors (multi-stakeholder, inter-governmental, private, government, civil society) to provide a high-level snapshot of the current state of AI ethics governance. Similarly to~\cite{floridi2018ai4people,jobin2019global}, no analysis of ethical AI implementations and principle-implementation gaps was conducted.

Hagendorff~\etal{}\cite{hagendorff2020ethics} evaluated 22 guidelines and found that diversity is lacking in the AI community, political abuse of AI systems (e.g., fake news, election fraud, automated propaganda) is also missed by the guidelines, and artificial general intelligence is not discussed. Our work extends~\cite{hagendorff2020ethics} with local and global analysis of AI principles (countries, continents), country AI readiness analysis,  principle-implementation gap analysis and mitigation.


\section{Study Methodology}~\label{methodology}

The \textbf{goal} of this study is to understand the landscape of AI principles around the world, 
through a contextual and holistic analysis of the characteristics of the different ethical AI principles proposed by countries from different continents, in order to identify potential 
principle-implementation gaps, their root causes, and formulate recommendations for gap mitigation. 
The \textbf{perspective} of this study is that of governments, national, and international organisations, which can leverage our findings to guide their AI strategy in developing and implementing AI principles adapted to their context. 
The \textbf{context} of this study is a set of 100 documents and websites containing 100 AI principles from 29 countries in the 6 continents (e.g., fairness, transparency) and 121 
implementations of these principles (e.g., guidelines, standards, tools) from 14 countries in the 6 continents, gathered online from reports of national/international organizations and websites. The data sources are selected based on their reliability, recency, and diversity. 
We address the following research questions:
\begin{quote}
	\textbf{RQ$_1$:} \textit{What are the characteristics of ethical AI principles across countries and continents?} This research question aims to help identify the countries (resp. continents) that released a high number of principles, the principles on which countries (resp. continents) place a greater emphasis (i.e., the most cited principles) and the principles shared across different countries (resp. continents). The common principles found between continents are compared to those from existing international organisations (i.e., OECD, GPAI, G20) to identify missing in the existing principles.
\end{quote}
\begin{quote}
	\textbf{RQ$_2$:} \textit{What is the current level of AI readiness in countries as well as the current implementations of the ethical AI principles?} 
	This research question seeks to identify the level of preparedness of countries based on scoring indexes~\cite{readiness_score2020,readiness_score2021} to implement the necessary changes required by the adoption and development of AI technologies. We also aim to understand how ethical AI principles are implemented in these countries. 
\end{quote}
\begin{quote}
	\textbf{RQ$_3$:} \textit{What are the gaps between ethical AI principles and their implementations as well as their root causes?} This research question aims to identify any gap that may exist between the stated AI principles and their implementation. We also aim to identify the root causes of these gaps, in order to formulate recommendations for their mitigation. 
\end{quote}
\begin{quote}
	\textbf{RQ$_4$:} \textit{What are potential solutions for principle-implementation gap mitigation?} This research question aims to propose some recommendations, to mitigate gaps between ethical AI principles and their implementation. 
\end{quote}

\begin{table*}[]
\centering
\caption{A record example of principles from all continents}
\label{sample-extracted}
\resizebox{\textwidth}{!}{
\scriptsize
\begin{tabular}{|l|l|l|l|l|l|l|l|l|}
\hline
\textbf{Provider}                                                                              & \textbf{Title}                                                                                                                                                                       & Location                                                 & Number & Principles                                                                                                                                                                                                                                                                                                                                                                                                                                  & References                                                                                                                                                              & Date & Type                                                          & Continent     \\ \hline
\begin{tabular}[c]{@{}l@{}}The Institute for \\ Ethical AI \& Machine\\  Learning\end{tabular} & \begin{tabular}[c]{@{}l@{}}The Responsible \\ Machine Learning \\ Principles\end{tabular}                                                                                            & \begin{tabular}[c]{@{}l@{}}United\\ Kingdom\end{tabular} & 8      & \begin{tabular}[c]{@{}l@{}}Human augmentation, \\ Bias evaluation, \\ Explainability By \\ Justification, Reproducible \\ operations, \\ Displacement strategy, \\ Pratical Accuracy, Trust \\ by privacy, Data risk \\ awareness\end{tabular}                                                                                                                                                                                              & \begin{tabular}[c]{@{}l@{}}https://ethical.institute\\ /principles.htmtl\end{tabular}                                                                                   &      & \begin{tabular}[c]{@{}l@{}}Research\\  Institute\end{tabular} & Europe        \\ \hline
\begin{tabular}[c]{@{}l@{}}Universite de \\ Montreal\end{tabular}                              & \begin{tabular}[c]{@{}l@{}}Montréal Declaration\\ for Responsible AI\end{tabular}                                                                                                    & Canada                                                   & 10     & \begin{tabular}[c]{@{}l@{}}well-being, autonomy, \\ intimacy and privacy, \\ solidarity, democracy, \\ equity, inclusion, caution,\\  responsibility and \\ environmental sustainability\end{tabular}                                                                                                                                                                                                                                       & {\color[HTML]{000000} \begin{tabular}[c]{@{}l@{}}https://www.montreal\\ declaration-responsible\\ ai.com/reports-of-montreal\\ -declaration\end{tabular}}               & 2017 & University                                                    & North America \\ \hline
IEEE                                                                                           & \begin{tabular}[c]{@{}l@{}}Ethically Aligned Design:\\  A Vision for Prioritizing \\ Human Well-being with \\ Autonomous and Intelligent\\ Systems - General Principles\end{tabular} & US                                                       & 5      & \begin{tabular}[c]{@{}l@{}}Human Rights;Prioritizing \\ Well-being;Accountability;\\ Transparency;A/IS Technology \\ Misuse and Awareness of It\end{tabular}                                                                                                                                                                                                                                                                                & {\color[HTML]{000000} \begin{tabular}[c]{@{}l@{}}https://standards.ieee.org/\\ content/dam/ieee-standards/\\ standards/web/documents/\\ other/ead\_v2.pdf\end{tabular}} & 2018 & \begin{tabular}[c]{@{}l@{}}Inter\\ national\end{tabular}      & North America \\ \hline
India Government                                                                               & Ethics and Human Rights                                                                                                                                                              & India                                                    & 5      & \begin{tabular}[c]{@{}l@{}}equality, safety \& reliability, \\ inclusivity \& non-discrimination, \\ transparency, accountability and \\ privacy \& security\end{tabular}                                                                                                                                                                                                                                                                   & {\color[HTML]{000000} \begin{tabular}[c]{@{}l@{}}https://indiaai.gov.in/\\ research-reports/responsible\\ -ai-part-1-principles-for-\\ responsible-ai\end{tabular}}     & 2021 & Government                                                    & Asia          \\ \hline
AI Forum                                                                                       & Trustworthy AI in Aotearoa                                                                                                                                                           & \begin{tabular}[c]{@{}l@{}}New \\ Zealand\end{tabular}   & 5      & \begin{tabular}[c]{@{}l@{}}Fairness and Justice;Reliability, \\ Security and Privacy;Transparency\\ ;Human Oversight and \\ Accountability;Well being\end{tabular}                                                                                                                                                                                                                                                                          & {\color[HTML]{000000} \begin{tabular}[c]{@{}l@{}}https://data.govt.nz/assets/\\ data-ethics/algorithm/\\ Trustworthy-AI-in-\\ Aotearoa-March-2020.pdf\end{tabular}}     & 2020 & Government                                                    & Ocenia        \\ \hline
Research ICT Africa                                                                            & \begin{tabular}[c]{@{}l@{}}Recomendations on the \\ inclusion subSaharan Africa\\  in Global AI Ethics\end{tabular}                                                                  & \begin{tabular}[c]{@{}l@{}}South \\ Africa\end{tabular}  & 6      & \begin{tabular}[c]{@{}l@{}}Introduce safeguards to balance AI\\ opportunities and risks; Protect individual\\  and collective privacy rights in crossborder\\  data flows; Define African values for AI \\ and align AI frameworks with such values; \\ Practise fair and socially-responsible AI; \\ Build inclusive partnerships based on \\ community and cocreation; Adopt an \\ adaptive, open minded\\ and humble approach\end{tabular} & {\color[HTML]{000000} \begin{tabular}[c]{@{}l@{}}https://researchictafrica.net/\\ wp/wp-content/uploads/2020/\\ 11/RANITP2019-2-AI-Ethics.pdf\end{tabular}}             & 2019 & \begin{tabular}[c]{@{}l@{}}Research \\ Institute\end{tabular} & Africa        \\ \hline
\begin{tabular}[c]{@{}l@{}}Chile's Ministry \\ of Science\end{tabular}                         & \begin{tabular}[c]{@{}l@{}}Cross-cutting AI \\ principles\end{tabular}                                                                                                               & Chile                                                    & 4      & \begin{tabular}[c]{@{}l@{}}IA with a focus on people's well-being, \\ respect for human rights and security; \\ IA for sustainable development; \\ Inclusive AI; Globalized \\and evolving AI\end{tabular}                                                                                                                                                                                                                                    & {\color[HTML]{000000} \begin{tabular}[c]{@{}l@{}}https://www.carey.cl/en/\\ chile-presents-its-first-\\ national-artificial-intelligence\\ -policy/\end{tabular}}       & 2021 & Government                                                    & South America \\ \hline
\end{tabular}
}
\end{table*}

\subsection{Data collection}

Three selection criteria have been defined for data collection: reliability, 
recency, and diversity. 
The reliability criteria ensures that the data can be trusted or is provided by a trusted sources. More than 350 reports and websites from companies, national, and international organizations have been inspected to collect AI principles and proposed implementation tools. For each candidate report, we assessed the trustworthiness of the institutions behind the report. An institution is considered to be trustworthy if it is officially recognised by an official state (e.g., The White House) and an international organisation (e.g., UNESCO).

The recency criteria ensures that the data source contains recent information about ethical AI principles. From the 350 collected reports and websites, we have selected 100 relevant reports and websites published between 
2016 and 2022. To select a report, we first read the document/website and check if it contains a 
clear reference to ethical principles (e.g., \textit{Our principles}, \textit{We follow these principles}), or to the implementation of principles (e.g., law and standard references, reference links for ethical AI softwares, training courses on AI ethics). Reports that did not contain ethical AI principles nor implementations were ignored.

The diversity criteria is used to ensure that we collect data 
from different countries and continents. The final set of reports analyzed in this paper 
covers countries from Europe, North America, South America, Oceania, Africa, and Asia. The references of these reports can be found in our replication package \cite{tech-report-v1}. 



\begin{table}[h]
\centering
\caption{Number of AI ethics implementations}
\scriptsize
\begin{tabular}{|p{4.0cm}|p{3.5cm}|}
\hline
\textbf{AI Ethics Implementation} & \textbf{Count} \\ \hline
Guidelines & 26 \\ \hline
Regulations \& Laws & 20 \\ \hline
Standards & 8 \\ \hline
Governance & 7 \\ \hline
Checklists & 6 \\ \hline
Algorithmic Assessment & 16 \\ \hline
Software & 28 \\ \hline
Training & 10 \\ \hline
\textbf{Total} & \textbf{121} \\ \hline
\end{tabular}
\label{tab:aiethics}
\end{table}

\subsection{Data extraction}

In the hundred reports and websites, we have manually selected 100 ethical AI principles randomly, representing a confidence level of 90\% with an error margin of 8.25\%. Each principle contains 2 to 15 keywords including Safe, Safety, Transparent, Transparency, Robust, Robustness, Explainable, Explainability, Fair, Fairness, Secure, Security, Responsible, Responsibility, Inclusive, and Inclusiveness.

Using Microsoft Excel, we have built regular expressions to extract these string keywords. The following regular expression was used to extract each principle keyword: \textit{*Safe*} (Safety), \textit{*Transparen*} (Transparency), \textit{*Robust*} (Robustness), \textit{*Explainab} (Explainability), \textit{*Fair*} (Fairness), \textit{*Secur*} (Security), \textit{*Responsib*} (Responsibility), and \textit{*Inclusive*} (Inclusiveness). For each principle, we extracted and recorded the following information: 
the name of the report's provider (e.g., US Department of Defense, IEEE), the title of the report, the location where the report is published, the number of principles contained in the report, the contained principles, the reference to get the principles, the date when the principle was released, the type of institution (e.g., government, international, company, university), and continent. An example of recorded information 
is shown in Table~\ref{sample-extracted}. The complete dataset is available in our replication package \cite{tech-report-v1}. 

We have extracted 521 occurrences of AI principle keywords from 29 countries: Canada, United States (US), United Kingdom (UK), China, Australia, Singapore, United Arab Emirates (UAE), Spain, Switzerland, Chile, Colombia, Sweden, Belgium, Israel, Amsterdam, New Zeland, India, Germany, Hong Kong, South Korea, South Africa, Finland, Japan, France, Netherlands, Norway, Russia, Ireland, and Italy. Table~\ref{fig:nbrprinccoun} presents the distribution of these principles across the countries. The 521 occurrences are grouped using 33 distinct keywords: Fairness, Transparency, Accountability, Safety, Security, Robustness, Explainability, Interpretability, Well-being, Human Oversight, Human Rights, Sustainability, Equity, Reliability, Privacy, Justice, Autonomy, Human Dignity, Responsibility, Solidarity, Beneficence, Non-maleficence, Non-discrimination, Contestability, Human-centric, Inclusiveness, Trustworthy, Democracy, Governance, Bias, Integrity, Controllability, and Accuracy.   

We also extracted 121 AI ethics implementations 
from the hundred reports and websites (see Table~\ref{tab:aiethics}), representing a confidence level of 95\% with an error margin of 8.91\%. In Table~\ref{tab:aiethics}, the extracted AI ethics implementations are grouped into 8 categories: guidelines, regulations \& laws, standards, governance, checklists, assessment, tools, and training. Categories such as guidelines, regulations \& laws, standards, and governance are more abstract and they provide orientations and rules to follow for AI ethics implementation. Other categories like checklists, assessment, and tools are more technical and they provide practical solutions for AI ethics implementation. The last category (i.e., training) contains courses about communicating and learning AI ethics.



\subsection{Data Processing and Analysis}

We processed the extracted data 
using word frequency analysis.
The word frequency analysis is performed by counting the number of AI principle occurrences per country and per continent. The aim being to identify potential characteristics such as the countries (resp. continents) that released a high number of principles, the principles on which specific countries (resp. continents) place a greater emphasis and the principles shared across different countries (resp. continents). 
Then, findings for the defined research questions are presented through point clouds using RStudio Integrated Development Environment (IDE). Point clouds show small and large circles depending on the number of occurrences of the principle; thus helping to identify the principle emphasis in the country (resp. continent). 


\section{Evaluation of ethical AI principles}~\label{evaluation}

In this section, we analyze the processed data and provide answers to our first research question (i.e., RQ1). 
We organise our analysis in two parts: (1) we perform a \textit{contextual} analysis of ethical AI principles; by identifying and analyzing AI ethical principles defined for specific countries, and (2) We perform a 
\textit{holistic} analysis by considering ethical AI principles that 
are generic, i.e., principles stated at a global scale (e.g., continent-level).

\begin{table}[]
\centering
\caption{Number of principles per country}
\label{fig:nbrprinccoun}
\begin{tabular}{|p{3.8cm}|p{3.5cm}|}
\hline
\textbf{Country} & \textbf{Total Number of principles} \\ \hline
Canada           & 29                       \\ \hline
United States (US)              & 167                      \\ \hline
United Kingdom (UK)              & 53                       \\ \hline
China            & 27                       \\ \hline
Australia        & 12                       \\ \hline
Singapore        & 14                       \\ \hline
United Arab Emirates (UAE)              & 4                        \\ \hline
Spain            & 5                        \\ \hline
Switzerland      & 10                       \\ \hline
Chile            & 4                        \\ \hline
Colombia         & 10                       \\ \hline
Sweden           & 9                        \\ \hline
Belgium          & 12                       \\ \hline
Israel           & 6                        \\ \hline
Amsterdam        & 5                        \\ \hline
New Zeland       & 11                       \\ \hline
India            & 5                        \\ \hline
Germany          & 28                       \\ \hline
Hong Kong        & 3                        \\ \hline
South Korea      & 19                       \\ \hline
South Africa     & 6                        \\ \hline
Finland          & 10                       \\ \hline
Japan            & 25                       \\ \hline
France           & 14                       \\ \hline
Netherlands      & 3                        \\ \hline
Norway           & 7                        \\ \hline
Russia           & 10                       \\ \hline
Ireland          & 9                        \\ \hline
Italy            & 4                        \\ \hline
\textbf{Total} & \textbf{521} \\ \hline
\end{tabular}
\end{table}


\subsection{Contextual evaluation}
To identify the characteristics of ethical AI
principles across countries (RQ1), 
we have formulated two sub-research questions: \textit{What are the countries that released a high number of principles ?} and \textit{What are the most cited principles in the countries as well as the principles shared across different countries ?}

\begin{figure*}[]
\includegraphics[scale=0.73]{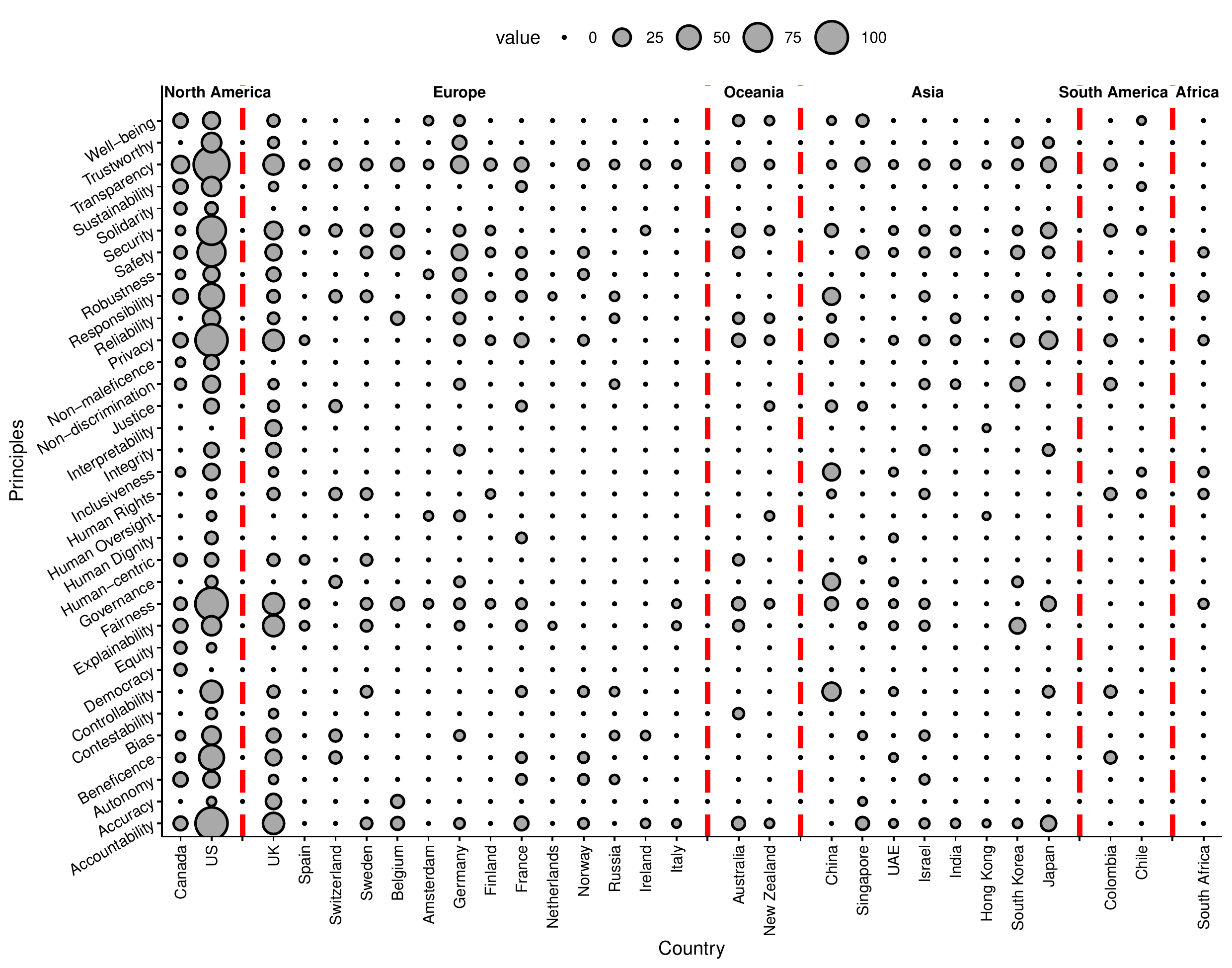}
\caption{Number of occurrences per principle per country}
\label{fig:nbrprinccounall}
\end{figure*}

\subsubsection{What are the countries that released a high number of principles ?} The aim of this question is to identify countries which are more involved in releasing national ethical AI principles. In Table~\ref{fig:nbrprinccoun}, United States (US) have the highest number of principles among the 29 countries. The second is United Kingdom (UK). Canada and Germany are respectively in third and fourth position among the 29 countries. Next come China and Japan in fifth and sixth position, respectively, among the 29 countries. In North America, US has released the highest number of principles followed by Canada. Countries like Mexico are not mentioned because they have not yet released any national ethical AI principles. However, Some of the missing countries are part of (or in the process of becoming member of) international organizations such as the Organization for Economic Cooperation and Development (OECD), which will be described later. 

In South America, Colombia is the first country to release a national ethical AI principles, followed by Chile. Other South American countries such as Argentina and Brazil do not have national ethical AI principles, but they are in a partnership with OECD. Chile, Costa Rica, Colombia, and Mexico are also OECD members. In Europe, UK is the first country to publish national ethical AI principles, followed by Germany. France appears in the third position followed by Belgium. In Asia, China is the first country with the highest number of principles and Japan occupies the second position. Next come South Korea and Singapore in third and fourth position, respectively. In Oceania, Australia is the first country with the highest number of principles followed by New Zealand. Other countries such as Nauru did not release national principles. In Africa, South Africa is the first to have national principles via the Research ICT Africa (RIA)\footnote{https://researchictafrica.net/}. Other African's countries do not yet have national ethical AI principles, but some are part of (or in the process of becoming member of) OECD.


\subsubsection{What are the most cited principles in the countries as well as the principles shared across different countries?} 

This research question aims to identify the principles on which countries place greater emphasis (i.e., the most cited principles) and the principles shared across different countries. 
Fig.~\ref{fig:nbrprinccounall} shows the number of principle occurrences per country from the set of principles extracted from the selected reports. A high (resp. low) circle diameter means a high (resp. low) number of principle occurrences in the country.

In North America, the Transparency principle has the largest frequency in United States and Canada. Transparency and Privacy principles are both cited in United States and Canada. In United States, we have identified 5 
top frequent principles as follows:  Transparent, Fairness, Accountability, Privacy, and Security. In Canada, the most frequent 
principles are Transparency, Responsibility, Privacy, Sustainability, Autonomy, and Well-being.

In Europe, the Transparency principle is the most frequent principle across 13 countries: UK, Germany, France, Belgium, Switzerland, Finland, Sweden, Norway, Spain, Amsterdam, Russia, Ireland, and Italy. It is described in Fig.~\ref{fig:nbrprinccounall} by gray circle appearing in these countries. The second most frequently occurring principle is Fairness. This principle covers 9 countries: UK, Belgium, Sweden, Germany, France, Spain, Amsterdam, Finland, and Italy. Security, Responsibility and Accountability principles are found in 8 countries (see Fig.~\ref{fig:nbrprinccounall}). The Security principle appears in Canada, US, UK, Spain, Switzerland, Sweden, Belgium, Germany, Finland, and Ireland. The Responsibility principle is found in Germany, UK, Switzerland, Sweden, France, Finland, Russia, and Netherlands. The Accountability principle appears in UK, France, Belgium, Sweden, Germany, Norway, Ireland, and Italy. Explainability and Safety principles appear in 7 countries including UK, Germany, Sweden, and France. The Privacy principle appears in 6 countries including UK, France, Norway, and Germany.


\begin{table}[]
\centering
\caption{Number of principles per continent}
\label{fig:nbrprinccont}
\begin{tabular}{|p{3.8cm}|p{3.5cm}|}
\hline
\textbf{Continent} & \textbf{Total Number of principles} \\ \hline
Europe          & 210    \\ \hline
Asia              & 103  \\ \hline
North America            & 201
                       \\ \hline
South America          & 14                     \\ \hline
Africa        & 6                      \\ \hline
Oceania        & 23                       \\ \hline
\end{tabular}
\end{table}

In Asia, the Transparency principle is the most frequent principle across 8 countries: Japan, Singapore, South Korea, Israel, India, China, United Arab Emirates (UAE), and Hong Kong (see Fig.~\ref{fig:nbrprinccounall}). The second most frequent principle is Accountability. This principle is found in 7 countries: Japan, Singapore, Israel, South Korea, India, UAE, and Hong Kong. Safety, Security, and Privacy principles appeared in 
6 countries. The Safety principle appears in South Korea, Singapore, Japan, Israel, India, and UAE; Security and Privacy principles appear in Japan, China, Israel, South Korea, India, and UAE. Other principles like Fairness, Explainability, and Responsibility are ranked in fourth position and found in 4 countries. The Fairness principle appears in China, Singapore, Israel, and UAE; and the Explainability principle appears in South Korea, Israel, UAE, and Singapore. The Responsibility principle is found in China, Japan, South Korea, and Israel.

\begin{boxblock}{Summary 1}
    \begin{itemize}
      \item In the studied samples, United States, United Kingdom, and Canada released the highest number of  principles between 2016 and 2021.
      \item Most countries from North America, Europe, Asia, Oceania, and South America placed a greater emphasis on the Transparency principle, identified by a high number of occurrences.  
      \item In North America, Transparency and confidentiality are the principles most frequently cited by countries.
      \item Transparency, Fairness, and Security are the principles most frequently stated by countries in Europe.
      \item In Asia, Transparency and Accountability  are the most frequently stated principles (by countries). 
      \item Transparency, Fairness, Accountability,  Security, Privacy, Reliability, and Well-being are the most frequently cited principles, by the countries in Oceania.
      \item In South America, Colombia and Chile share principles such as Security (resp. Safety), Human Rights (resp. Human-centric), and Non-discrimination (resp. Inclusiveness). 
      \item In Africa, South Africa is focused on Fairness, Inclusiveness, Safety, Privacy, Responsibility, and Human Rights principles.
    \end{itemize}
\end{boxblock}

Oceania, Australia and New Zealand have 7 principles in common: Transparency, Fairness, Accountability, Security, Privacy, Reliability, and Well-being. In addition, Australia has Explainability, Contestability, and Human-centric principles whereas New Zealand has Human Oversight and Justice principles. In South America, Columbia and Chile follows different principles. Columbia follows the Transparency principle and 7 others principles including Security, Privacy, Responsibility, Non-discrimination, Controllability, and Beneficence. Chile has the following principles: Human-centric, Sustainability, Safety, Inclusiveness, and Democracy. In Africa, South Africa follows Fairness, Inclusiveness, Safety, Privacy, Responsibility, and Human Rights principles. 







\subsection{Holistic evaluation}
In this section, ethical AI principles are analyzed on a global scale to identify the principles on which continents place a greater emphasis (i.e. the most cited principles) and the principles shared across different continents. The identified shared principles are compared with common principles provided by international organizations such as OECD, GPAI, and G20 to which different countries are affiliated. To structure our analysis, we have defined two sub-research questions: 
\textit{What are the most cited principles in the continents as well as the principles shared across different continents?} and \textit{How do continent-specific principles differ from those provided by international organizations (i.e., OCECD, GPAI or G20)?}. 



\subsubsection{What are the most cited principles in the continents as well as the principles shared across different continents ?}
\begin{figure}[h]
\centering
\includegraphics[scale=0.61]{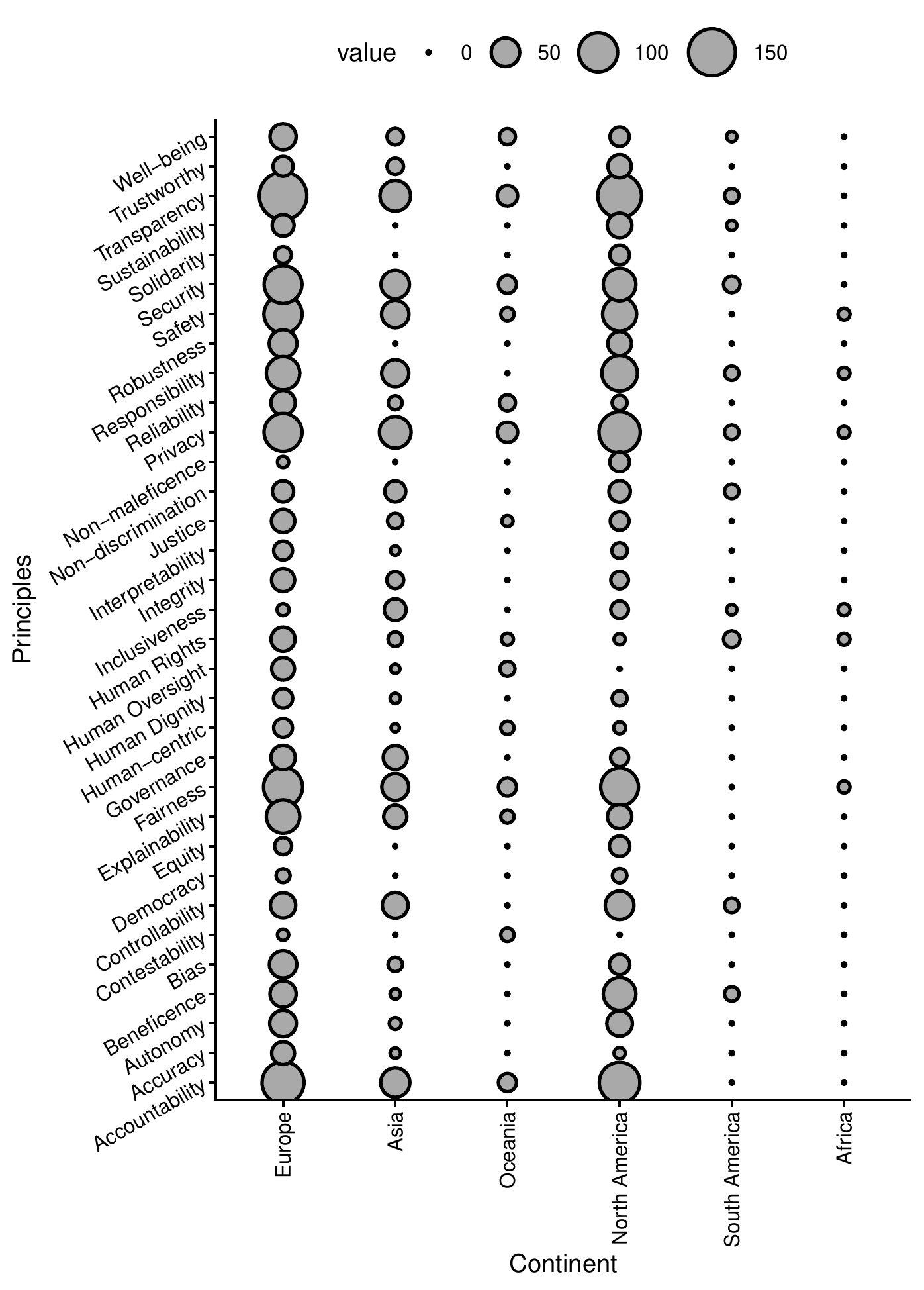}
\caption{Number of occurrences per principle per continent}
\label{fig:nbrprinccont1}
\end{figure}

This research question aims to identify the principles on which continents place greater emphasis (i.e., the most cited principles) and the principles shared across different continents.  Fig.~\ref{fig:nbrprinccont1} shows that Europe and North America have the 
highest number of ethical AI principles. Human Rights and Privacy principles are shared by all continents. Transparency, Fairness, Security, and Safety principles have the highest number of occurrences and they appear in 5 continents: Europe, North America, Asia, Oceania, South America, and Africa. Responsibility, Well-being, and Inclusiveness principles are also found in 5 continents: Europe, North America, Asia, South America, Oceania, and Africa. Accountability and Explainability principles appear in 4 continents with a high number of occurrences: Europe, North America, Asia, and Oceania. 

\textit{Global principles.} We 
identified 11 global principles that cover most continents with a high number of occurrences: Transparency, Privacy, Fairness, Security, Safety, Responsibility, Accountability, Explainability, Well-being, Human Rights, and Inclusiveness.


\subsubsection{How do continent-specific principles differ from those provided by international organizations such as OCECD, GPAI or G20 ?} In this research question, we aim to examine the alignment between the 
global principles developed on different continents and those proposed by international organizations (i.e., OECD, GPAI, G20).

\paragraph{GPAI}

GPAI\footnote{https://gpai.ai/} is a multi-stakeholder association consisting of scientists, industries, civil society, governments, international organisations and academia aiming to bridge the gap between theory and practice on AI. They released 5 principles: (1) Inclusive growth, sustainable development and well-being; (2) Human-centred values and fairness; (3) Transparency and explainability; (4) Robustness, security, and safety; and (5) Accountability. The 11 global principles identified in the previous question are similar to these principles. Well-being, Human Rights, Inclusiveness, and Fairness principles cover principles (1) and (2), except the Sustainability principle. Security, Safety, Transparency, Explainability, and Accountability principles cover principles (3), (4) and (5), except the Robustness principle. The Responsibility principle encompasses Sustainability, Robustness principles and several other principles (e.g., trustworthiness, reliability, impartiality). GPAI principles do not contain the Responsibility principle which is important (it appears in 5 continents with a high number of occurrences).

\paragraph{OECD}

OECD\footnote{https://oecd.ai/en/} is an international organization for economic studies with country members from all continents. OECD promotes 5 AI Principles: (1) Inclusive growth, sustainable development and well-being; (2) Human-centered values and fairness; (3) Transparency and explainability; (4) Robustness, security and safety; and (5) Accountability. From the identified global principles, (1) is covered by Inclusiveness, Well-being, and Responsibility principles. Principle (2) is covered by Human Rights and Fairness. (3) and (4) are covered by Transparency, Security, Safety, Responsibility, and Explainability principles. Principle (5) is also found in the global principles. OECD AI principles do not contain Privacy and Responsibility principles; yet they appeared in 5 continents with a high number of occurrences. 

\paragraph{G20}

G20\footnote{https://g20.org/} is an intergovernmental forum with 19 country members and the European Union, that addresses global economy issues (e.g., climate change). G20 AI principles are drawn from the OECD AI principles. Thus, G20 AI principles do not also contain Privacy and Responsibility principles. 

\begin{boxblock}{Summary 2}
    \begin{itemize}
      \item According to our analysis, 
      North America and Europe released the highest number of principles between 2016 and 2021.
      \item The 11 global principles that cover most continents with a high number of occurrences are: Privacy, Transparency, Fairness, Security, Safety, Responsibility, Accountability, Explainability, Well-being, Human Rights, and Inclusiveness.
      \item GPAI ethical AI principles do not include the 
      Responsibility principle. 
      \item OECD and G20 ethical AI principles do not include Privacy and Responsibility principles. 
    \end{itemize}
\end{boxblock}

\begin{table*}[]
\parbox{.48\linewidth}{
\centering
\caption{country AI Readiness score (2021)}
\label{readscorecoun}
\begin{tabular}{|l|l|l|l|}
\hline
\textbf{Country} & \textbf{Government} & \textbf{Technology} & \textbf{Data\& Infra.} \\ \hline
US & \cellcolor[HTML]{DCDCDC}88.46 & \cellcolor[HTML]{DCDCDC}83.31 & \cellcolor[HTML]{DCDCDC}92.71 \\ \hline
Singapore & \cellcolor[HTML]{DCDCDC}94.88 & \cellcolor[HTML]{DCDCDC}66.69 & 85.8 \\ \hline
UK & \cellcolor[HTML]{DCDCDC}85.69 & \cellcolor[HTML]{DCDCDC}67.26 & \cellcolor[HTML]{DCDCDC}90.81 \\ \hline
Finland & \cellcolor[HTML]{DCDCDC}88.45 & 63.85 & 85.4 \\ \hline
Netherlands & 80.42 & \cellcolor[HTML]{DCDCDC}66.17 & \cellcolor[HTML]{DCDCDC}88.92 \\ \hline
Sweden & 80.76 & \cellcolor[HTML]{DCDCDC}67.37 & 86.36 \\ \hline
Canada & \cellcolor[HTML]{DCDCDC}84.36 & 63.75 & 85.08 \\ \hline
Germany & 78.04 & \cellcolor[HTML]{DCDCDC}67.68 & 86.07 \\ \hline
Denmark & 83.5 & 63.24 & 84.14 \\ \hline
Republic of Korea & \cellcolor[HTML]{DCDCDC}85.27 & 58.49 & 85.89 \\ \hline
France & 82.1 & 60.61 & \cellcolor[HTML]{DCDCDC}86.53 \\ \hline
Japan & 81.9 & 59.31 & \cellcolor[HTML]{DCDCDC}87.32 \\ \hline
Norway & 84.24 & 59.25 & 84.91 \\ \hline
Australia & 83.79 & 57.07 & 85.37 \\ \hline
China & 83.79 & 61.33 & 78.15 \\ \hline
Luxembourg & 82.67 & 50.66 & \cellcolor[HTML]{DCDCDC}86.8 \\ \hline
Ireland & 74.7 & 61.11 & 82.59 \\ \hline
Taiwan & 77.59 & 59.42 & 78.92 \\ \hline
UAE & 79.41 & 53.33 & 82.05 \\ \hline
Israel & 64.64 & 65.87 & 79.52 \\ \hline
\end{tabular}
}
\hfill
\parbox{.48\linewidth}{
\centering
\caption{country AI Readiness score per year (2020-2021)}
\label{readscorecounyear}
\begin{tabular}{|l|p{1.2cm}|p{1.2cm}|p{2.0cm}|}
\hline
\textbf{Country} & \textbf{2020} & \textbf{2021} & \textbf{Growth}\\ \hline
US & \cellcolor[HTML]{DCDCDC}85.479 & \cellcolor[HTML]{DCDCDC}88.16 & \cellcolor[HTML]{DCDCDC}2.681 \\ \hline
Singapore & \cellcolor[HTML]{DCDCDC}78.704 & \cellcolor[HTML]{DCDCDC}82.46 & \cellcolor[HTML]{DCDCDC}3.756 \\ \hline
UK & \cellcolor[HTML]{DCDCDC}81.124 & \cellcolor[HTML]{DCDCDC}81.25 & \cellcolor[HTML]{DCDCDC}0.126 \\ \hline
Finland & \cellcolor[HTML]{DCDCDC}79.238 & \cellcolor[HTML]{DCDCDC}79.23 & -0.008 \\ \hline
Netherlands & 75.297 & \cellcolor[HTML]{DCDCDC}78.51 & \cellcolor[HTML]{DCDCDC}3.213 \\ \hline
Sweden & \cellcolor[HTML]{DCDCDC}78.772 & \cellcolor[HTML]{DCDCDC}78.16 & -0.612 \\ \hline
Canada & 73.158 & 77.73 & \cellcolor[HTML]{DCDCDC}4.572 \\ \hline
Germany & \cellcolor[HTML]{DCDCDC}78.974 & 77.26 & -1.714 \\ \hline
Denmark & 75.618 & 76.96 & \cellcolor[HTML]{DCDCDC}1.342 \\ \hline
Republic of Korea & 77.695 & 76.55 & -1.145 \\ \hline
France & 73.767 & 76.41 & \cellcolor[HTML]{DCDCDC}2.643 \\ \hline
Japan & 73.303 & 76.18 & \cellcolor[HTML]{DCDCDC}2.877 \\ \hline
Norway & 74.43 & 76.14 & \cellcolor[HTML]{DCDCDC}1.71 \\ \hline
Australia & 73.577 & 75.41 & \cellcolor[HTML]{DCDCDC}1.833 \\ \hline
China & 69.08 & 74.42 & \cellcolor[HTML]{DCDCDC}5.34 \\ \hline
Luxembourg & 72.616 & 73.37 & \cellcolor[HTML]{DCDCDC}0.754 \\ \hline
UAE & 72.395 & 71.6 & -0.795 \\ \hline
Israel & 68.825 & 70.01 & \cellcolor[HTML]{DCDCDC}1.185 \\ \hline
\end{tabular}
}
\end{table*}

\section{Evaluation of ethical principle implementations}~\label{analysis}

In order to identify gaps between ethical AI principles and their implementations, it is important to review and understand the current implementations in the countries. In~\cite{readiness_score2020,readiness_score2021}, Oxford Insights and International Development Research Centre (IRDC) provide information about the AI readiness of countries in terms of governance, infrastructure, and technology for AI implementation support. Based on the information contained in ~\cite{readiness_score2020,readiness_score2021}, we perform a country AI readiness analysis between 2020 and 2021, to identify the current progression state of AI implementations in the different countries. Next, we perform an in-depth analysis of 
recent ethical AI implementations documents, such as guidelines, code of conduct (i.e., regulation and laws, standards), governance rules, ethical AI assessment tools (i.e., checklists, algorithmic assessments), implementation tools, and awareness activities (i.e., online training, campaigns). In the following, we report about these analysis; presenting  
the level of preparedness of different countries (i.e., their AI readiness) and the existing ethical AI implementations (RQ2). We also report about gaps identified between published ethical AI principles and their implementation (RQ3).

\subsection{What is the current level of AI readiness in countries?}

In order to identify the country AI readiness (RQ2),  scoring indexes~\cite{readiness_score2020,readiness_score2021} 
evaluates the current level of AI preparedness following different levels: (1) government, (2) technology factor, and (3) data and infrastructure. At the government level, the score gives insights about the willingness of the government of the country to adopt AI, and its ability to adapt and innovate. In the technology sector, the score gauges the level of supply of AI tools in the country.
At the data and infrastructure level, the score provides insights about the implementation of AI tools and assessing whether they are trained on high quality and representative data, and whether the government of the country has the appropriate infrastructure for delivering AI solutions and fostering its adoption by citizens~\cite{readiness_score2021}. 


Table~\ref{readscorecoun} shows the AI readiness score for 20 countries in 2021: US, Singapore, UK, Finland, Netherlands, Sweden, Canada, Germany, Denmark, Republic of Korea, France, Japan, Norway, Australia, China, Luxembourg, Ireland, Taiwan, UAE, and Israel. At the government level, Singapore comes in first position (94.88) followed by US (88.46). Finland is ranked in the third position (88.45) followed by UK (85.69). The Republic of Korea and Canada are respectively in fifth (85.27) and sixth (84.36) position. In the technology sector, US leads with a score of 83.31, followed by Germany with a score of 67.68. Next is Sweden, ranked in third position (67.37), followed by UK (67.26). Singapore and Netherlands are respectively in fifth (66.69) and sixth position (66.17). At the data and infrastructure level, US is ranked first (92.71) followed by UK (90.81). Netherlands is ranked in third position (88.92) followed by Japan (87.32). Next is Luxembourg and France appearing respectively in fifth (86.8) and sixth position (86.53). From these results, US, UK, and Germany are more favorable for AI development, investment, business, and research since they are well prepared in terms of AI technology, data and infrastructure, and AI governance. Other countries such as Finland, Sweden, Netherlands, Japan, and Singapore are also propice for investment, business, and research in AI. 


Table~\ref{readscorecounyear} shows the yearly AI readiness score of 18 countries: US, Singapore, UK, Finland, Netherlands, Sweden, Canada, Germany, Denmark, Republic of Korea, France, Japan, Norway, Australia, China, Luxembourg, UAE, and Israel. Overall, US and UK respectively maintained the first position (+2.681 increase) and second position (+0.126 increase) from 2020 to 2021. Singapore moved from the sixth position in 2020 to the second position in 2021 with a +3.756 increase. Finland moved from the third position in 2020 to the fourth position in 2021 with a -0.008 decrease. Similar to Finland, Germany's readiness score experienced a significant decrease of -1.714 from 2020 to 2021. Sweden's AI readiness score also decreased from 78.772 in 2020 to 78.16 in 2021 (-0.612 decrease). Netherlands have a +3.213 increase from 2020 to 2021. Canada have experienced a high growth from 2020 to 2021 with a +4.572 increase. China have the highest growth with a +5.34 increase from 2020 to 2021. Other countries like France, Japan, Norway, Australia, Israel, and Luxembourg also experienced a significant growth (as shown in Table~\ref{readscorecounyear}). From these results, China, Canada, Singapore, and Netherlands have a high potential of increase in the next years in terms of AI governance, technology, data and infrastructure. This also makes them favorable for investment, business, and research in AI.

\begin{boxblock}{Summary 3}
    \begin{itemize}
      \item At the government level, Singapore was the leader followed by US in 2021. In the technology sector, United States was the leader followed by Germany in 2021. At the data and infrastructure level, US was the leader followed by UK in 2021. Then, US, UK, and Germany are more favorable for investment, business, and research in AI.
      \item In terms of AI readiness, United States maintained the first position followed by United Kingdom from 2020 to 2021. China and Canada experienced the highest growth (resp. +5.34 and +4.572 increase) from 2020 to 2021. The AI readiness of Germany and Republic of Korea decreased significantly (resp. -1.714 and -1.145) from 2020 to 2021. Then, China and Canada have a high potential of increase in the next years, also making them favorable for investment, business, and research in AI.
    \end{itemize}
\end{boxblock}
\subsection{What are the current ethical AI principle implementations?}\label{existing-guidelines}

Several ethical AI guidelines, regulation, and tools have been proposed for designing, and implementing ethical AI systems. 
The ethical principle implementation materials found through our review (described in Sections~\ref{guidelines}-\ref{awareness}) are divided in 6 categories: guidelines, code of conduct (e.g., standards), governance tools, assessment tools (e.g., checklists), AI ethics softwares, and awareness materials (e.g., training courses). Guidelines, code of conduct, and governance tools provide high-level implementations of ethical AI principles while awareness materials, assessment tools, and softwares are more technical. 
Here after, the existing ethical AI implementation tools are presented (RQ2).

\begin{table*}[]
\caption{AI ethics guidelines}
\label{fig:aiethicsimpl}
\resizebox{\textwidth}{!}{
\begin{tabular}{|l|l|l|l|l|}
\hline
\textbf{Name} & \textbf{Author} & \textbf{Organization} & \textbf{Year} & \textbf{Location} \\ \hline
The Ethics of AI Ethics - An Evaluation of Guidelines & T. Hagendorff & \begin{tabular}[c]{@{}l@{}}University \\ of Tübingen\end{tabular} & 2020 & Germany \\ \hline
Understanding artificial intelligence ethics and safety & D. Leslie & \begin{tabular}[c]{@{}l@{}}The Alan Turing\\ Institute\end{tabular} & 2019 & US \\ \hline
The global landscape of AI ethics guidelines & A. Jobin et al. & ETH Zurich & 2019 & Switzerland \\ \hline
Ethics Guidelines for Trustworthy AI & AI HLEG & \begin{tabular}[c]{@{}l@{}}European \\ Commission\end{tabular} & 2019 & EU \\ \hline
\begin{tabular}[c]{@{}l@{}}From What to How: An Initial Review of Publicly \\ Available AI Ethics Tools, Methods and Research \\ to Translate Principles into Practices\end{tabular} & J. Morley et al. & \begin{tabular}[c]{@{}l@{}}Oxford Internet \\ Institute\end{tabular} & 2020 & UK \\ \hline
\begin{tabular}[c]{@{}l@{}}From Principles to Practice: An interdisciplinary \\ framework to operationalise AI ethics\end{tabular} & \begin{tabular}[c]{@{}l@{}}S. Hallensleben\\ C. Hustedt\end{tabular} & AIEI Group & 2020 & Germany \\ \hline
\begin{tabular}[c]{@{}l@{}}Building Ethics into AI: Lessons Learned from\\ Pioneers in the Trenches\end{tabular} & K. Baxter & \begin{tabular}[c]{@{}l@{}}Salesforce \\ Research\end{tabular} & 2019 & US \\ \hline
Technical and Organizational Best Practices &  & FBPML & 2021 & Netherlands \\ \hline
\begin{tabular}[c]{@{}l@{}}Advancing AI ethics beyond compliance From\\  principles to practice\end{tabular} & \begin{tabular}[c]{@{}l@{}}B. C. Goehring\\ et al.\end{tabular} & IBM & 2020 & US \\ \hline
Guidelines for AI Procurement &  & \begin{tabular}[c]{@{}l@{}}World Economic\\ Forum\end{tabular} & 2019 & Switzerland \\ \hline
Guidance on AI and data protection &  & \begin{tabular}[c]{@{}l@{}}Information \\ Commissioner Office\end{tabular} & 2020 & UK \\ \hline
Guidance on the AI auditing framework &  & \begin{tabular}[c]{@{}l@{}}Information \\ Commissioner Office\end{tabular} & 2020 & UK \\ \hline
\begin{tabular}[c]{@{}l@{}}Building Ethics into Privacy Frameworks\\  for Big Data and AI\end{tabular} &  & IAPP & 2018 & US \\ \hline
\begin{tabular}[c]{@{}l@{}}Ethical by Design: Ethics Best Practices \\ for Natural Language Processing\end{tabular} & J.L. Leidner & Thomson Reuters & 2017 & UK \\ \hline
Ethical OS Toolkit &  & \begin{tabular}[c]{@{}l@{}}IFTF \& Omidyar \\ Network\end{tabular} & 2018 & US \\ \hline
\begin{tabular}[c]{@{}l@{}}Data preprocessing techniques for \\ classification without discrimination\end{tabular} & F. Kamiran et al. & \begin{tabular}[c]{@{}l@{}}Eindhoven University \\ of Technology\end{tabular} & 2012 & Netherlands \\ \hline
Responsible Design Assistant &  & Responsible AI institute & 2020 & US \\ \hline
AI RFX Procurement Framework &  & \begin{tabular}[c]{@{}l@{}}Institute for Ethical \\ AI \& ML\end{tabular} & 2018 & UK \\ \hline
Ethically Aligned Design v1 and v2 &  & IEEE & 2017 & US \\ \hline
\begin{tabular}[c]{@{}l@{}}A Toolkit for Centering Racial Equity \\ Throughout Data Integration\end{tabular} & A. H. Nelson et al. & AISP & 2020 & US \\ \hline
Design Ethically Toolkit & Kat Zhou & University of Cambridge & 2016 & UK \\ \hline
\begin{tabular}[c]{@{}l@{}}Responsible AI – Key Themes, Concerns \& \\ Recommendations for European Research and \\ Innovation Summary of Consultation with \\ Multidisciplinary Experts\end{tabular} & S. Taylor et al. & HUB4NGI & 2018 & EU \\ \hline
Ethical Toolkit for Engineering/Design Practice & S. Vallor & Markkula Center & 2018 & US \\ \hline
\begin{tabular}[c]{@{}l@{}}Tools and Ethics for Applied Behavioural \\ Insights: The BASIC Toolkit\end{tabular} &  & OECD & 2019 & France \\ \hline
\begin{tabular}[c]{@{}l@{}}AI4People—An Ethical Framework for a Good\\ AI Society: Opportunities, Risks, Principles, \\ and Recommendations\end{tabular} & L. Floridi & Atomium-EISMD & 2018 & EU \\ \hline
Ethics Canvas &  & \begin{tabular}[c]{@{}l@{}}ADAPT Centre \& \\ Trinity College\end{tabular} & 2017 & Ireland \\ \hline
\end{tabular}
}
\end{table*}

\subsubsection{AI Ethics Guidelines}\label{guidelines}

Table~\ref{fig:aiethicsimpl} shows 26 ethical AI guidelines ranging from 2017 to 2022 and classified by name, author, organization, year, and location. 
T. Hagendorff~\cite{hagendorff2020ethics} analyzed 22 guidelines
and found omissions in these guidelines. They were not included in the 26 guidelines since the paper~\cite{hagendorff2020ethics} itself provides a summary of those guidelines. T. Hagendorff found that artificial general intelligence was not discussed due to the fact that the guidelines were mostly written by research groups with computer science background. He also observed that diversity among people and technology 
is lacking in the AI community. He also found that very few guidelines was provided to avoid AI systems being used for political abuses (e.g., fake news, election fraud, automated propaganda).

J. Morley \etal{}~\cite{morley2021initial} built a typology to apply ethics at each stage of the Machine Learning (ML) development pipeline by identifying the tools and methods available, and developers or companies researching and producing them. B.C. Goehring \etal{}~\cite{goehring2020advancing} proposes to educate students and professionals by adding AI ethics in business and law schools, computer science programs, technical organizations; setting guidelines and standards at national, supranational, and global levels; and creating an unified approach to ensure a robust, complete, and holistic assessment of present and future AI ethics implications. 

A. Kaminski \etal{}~\cite{kaminski2020principles} proposes to apply the Values, Criteria, Indicators, and Observables (VCIO) model to AI ethics. The ethical values defined include Transparency. For transparency, the criteria as well as indicators include the disclosure of the origin of the datasets (e.g., an indicator can be whether data is documented) and the disclosure of properties of algorithm/model used (e.g., an indicator can be whether the model has been tested/used before). Then, observables can be a comprehensible logging of all training and operating data, version control of datasets, and wide usage and testing of the model both in theory and practice.

D. Leslie~\cite{leslie2019understanding} describes constraints that an AI project must follow and explain how to practically build an ethical platform. An AI project must be ethically permissible, fair and non-discriminatory, worthy of public trust, and justifiable. Their proposed ethical platform has three building blocks: (1) Support, Underwrite, and Motivate (SUM) ethical values which are based on the notions of Respect, Connect, Care, and Protect; (2) Fairness, Accountability, Sustainability, and Transparency (FAST) actionable principles which provides the moral and practical tools to deliver safe and reliable AI solutions; and a (3) process-based governance (PBG) framework to operationalize the SUM Values and the FAST Track Principles across the entire AI project delivery workflow.

A.H. Nelson \etal{}~\cite{nelson2020toolkit} recommands to normalize, organize, and operationalize racial equity throughout data integration. They suggest an ongoing process at each stage of the data life cycle-planning, i.e., data collection (e.g., shared data, inclusive data), data access (open data, restricted data, available data), use of algorithms and statistical tools (i.e., following Responsibility, Explainability, Accuracy, and Auditability, Fairness principles), data analysis (e.g., using participatory research to bring multiple perspectives to the interpretation of the data), reporting and dissemination (e.g.,  providing clear documentation of the data analysis process with analytic files for reproducibility).

The AI High Level Expert Group (HLEG)~\cite{smuha2019ethics} proposed technical and non-technical methods to implement AI ethics. Technical methods include architectures for Trustworthy AI, ethics and rules of law by design (e.g., transparency-by-design, fairness-by-design), explanation methods (e.g., IBM XAI), testing and validation (e.g., red teaming, bug bounty programs). Non technical-methods include regulation, codes of conduct, standardization, certification, accountability via governance frameworks, diversity and inclusive design teams, education and awareness to foster an ethical mind-set, stakeholder participation and social dialogue.

The Foundation for Best Practices in Machine Learning (FBPML)~\cite{fbpml_doc} provides technical and organisational best practices. These organisational best practices include a clear definition and diversity of team roles, determination and definition of an appropriate, feasibility, and solvability of the business problem, determination of the most desirable and feasible model to achieve the desired product outcomes, management and monitoring of product pipeline during the product management lifecycle (i.e., Design, Exploration, Experimentation, Development, Implementation, Operationalisation). Technical best practices include the use of fairness and non-discrimination in products and models, data quality control, representativeness of the product model and data, performance and robustness of model outcomes, monitoring and maintenance of products and models, and explainability of model functions and outputs.

The World Economic Forum (WEF)~\cite{wef-report} proposes data governance guidelines including the use of procurement processes with focus on outlining problems and opportunities, the definition of the public benefit of using AI while assessing risks, inclusion of procurement's within a strategy for AI adoption across government, highlighting the technical and ethical limitations of using the data to avoid bias issues (e.g., data quality such as data completeness, representativenes and accuracy), working with a diverse and multidisciplinary team, focusing throughout the procurement
process on mechanisms of accountability and transparency norms. Other ethical AI guidelines were also proposed by international organizations such as IEEE~\cite{ethics_ieee_2020}, OECD~\cite{/content/publication/9ea76a8f-en}, International Association of Privacy Professionals (IAPP)~\cite{iapp_report}, Atomium-European Institute for Science, Media and Democracy (EISMD)~\cite{pagallo2019good}, Hub for Next Generation Internet (HUB4NGI) Consortium~\cite{taylor2018responsible}; national organizations such as the UK's Information Commissioner Office~\cite{uk_ico_report, uk_ico_report1}; companies such as Thomson Reuters~\cite{leidner2017ethical} and Markkula Center~\cite{vallor2018ethical}; and universities such as the University of Cambridge~\cite{kat_zhou}, Trinity College~\cite{adapt_centre}, and ETH Zurich~\cite{jobin2019global}.

\begin{table}[]
\caption{AI ethics Regulation \& Laws}
\label{fig:aiethicsregulation}
\centering
\begin{tabular}{|l|l|l|}
\hline
\textbf{Name} & \textbf{Country} & \textbf{Year} \\ \hline
Privacy Act & Canada & 1985 \\ \hline
\begin{tabular}[c]{@{}l@{}}The Personal Information Protection \\ and Electronic Documents Act (PIPEDA)\end{tabular} & Canada & 2000 \\ \hline
Financial Consumer Agency of Canada Act & Canada & 2001 \\ \hline
\begin{tabular}[c]{@{}l@{}}Health Insurance Portability and \\ Accountability Act (HIPAA)\end{tabular} & US & 1996 \\ \hline
Privacy Act & US & 1974 \\ \hline
Privacy Protection Act & US & 1980 \\ \hline
California Consumer Privacy Act (CCPA) & US & 2018 \\ \hline
EU-U.S. and Swiss-U.S. Privacy Shield & US & 2016 \\ \hline
Gramm–Leach–Bliley Act & US & 1999 \\ \hline
Fair Credit Reporting Act & US & 2018 \\ \hline
Data Protection Act & UK & 2018 \\ \hline
General Data Protection Regulation (GDPR) & EU & 2016 \\ \hline
Privacy Act & Australia & 1988 \\ \hline
Personal Information Protection Law (PIPL) & China & 2021 \\ \hline
Cyber Security Law (CSL) & China & 2016 \\ \hline
Personal Data Protection Act & Singapore & 2012 \\ \hline
\begin{tabular}[c]{@{}l@{}}Protection from Online Falsehoods and \\ Manipulation Act (POFMA)\end{tabular} & Singapore & 2019 \\ \hline
Protection of Personal Information Act & \begin{tabular}[c]{@{}l@{}}South \\Africa\end{tabular} & 2013 \\ \hline
Data Protection Law & Dubai & 2021 \\ \hline
Bundesdatenschutzgesetz (BDSG) & Germany & 2009 \\ \hline
\end{tabular}
\end{table}

\subsubsection{AI Ethics Code of Conduct}\label{code-conduct}

The code of conduct describes regulatory norms, laws, and standards defined by governments and international organizations. 

\paragraph{\textbf{Regulations \& Laws}}

Table~\ref{fig:aiethicsregulation} shows 20 ethical AI regulations and laws ranging from 1974 to 2018 and classified by name, country, and year of publication. In Canada, AI ethics is supported by the protection of personal information laws such as the Privacy Act (1985)~\cite{can_priv}, 
the Personal Information Protection and Electronic Documents Act (2000)~\cite{pipeda}, and the Financial Consumer Agency of Canada Act (2001)~\cite{finac}. In US, healthcare data and personal information is protected by the Health Insurance Portability and Accountability Act of 1996 (HIPAA)~\cite{hippa}. The protection of personal information laws include Privacy Act (1974)~\cite{us_priv}, Privacy Protection Act (1980)~\cite{us_protect}, and California Consumer Privacy Act (2018)~\cite{cal_act}. The EU-U.S. and Swiss-U.S. Privacy Shield (2016)~\cite{eu_us_act} law protects personal data transfer between EU and US. US Financial institutions have to comply Fair Credit Reporting Act (2018)~\cite{credit_act} and Gramm–Leach–Bliley Act (1999)~\cite{glb_act} for the collection, dissemination, and use of consumer information (e.g., credit reports). Other laws were proposed by UK (i.e., Data Protection Act~\cite{uk_act}), EU (i.e., General Data Protection Regulation~\cite{gdpr_act}), Australia (i.e., Privacy Act~\cite{aus_act}), China (i.e., Personal Information Protection Law~\cite{china_act}, Cyber Security Law~\cite{china1_act}), Singapore (i.e., Personal Data Protection Act~\cite{sg_act}, Protection from Online Falsehoods and Manipulation Act~\cite{sg1_act}), South Africa (i.e., Protection of Personal Information Act~\cite{sa_act}), Dubai (i.e., Data Protection Law~\cite{du_act}), and Germany (i.e., Bundesdatenschutzgesetz~\cite{ger_act}).

\begin{table*}[]
\centering
\caption{AI ethics Standards}
\label{fig:aiethicsstandards}
\begin{tabular}{|l|l|l|}
\hline
\textbf{Name} & \textbf{Organization} & \textbf{Year} \\ \hline
Global AI Standards Repository & OCEANIS & 2019-current \\ \hline
Ethical design and use of automated decision systems (CAN/CIOSC 101:2019) & CIO Strategy Council & 2019-present \\ \hline
ISO/IEC JTC 1/SC 42 Standards for Artificial Intelligence & ISO & 2017-current \\ \hline
Artificial Intelligence and User Trust (NISTIR 8332)& NIST & 2021 \\ \hline
Artificial Intelligence and Public Standards & Committee on Standards in Public Life UK & 2020 \\ \hline
The IEEE Global Initiative on Ethics of Autonomous and Intelligent Systems & IEEE SA & 2019 \\ \hline
Code of Ethics and Professional Conduct & ACM & 2018 \\ \hline
Code of Ethics and Professional Conduct & Australian Computer Society &  \\ \hline
\end{tabular}%
\end{table*}

\paragraph{\textbf{Standards}}

Table~\ref{fig:aiethicsstandards} presents 8 ethical AI standards ranging from 2017 to now and classified by name, organization, and year of publication. The Open Community for Ethics of Autonomous and Intelligent Systems (OCEANIS)~\cite{oceanis} is a global AI standards repository that captures AI and Autonomous and Intelligent Systems standards 
from different organizations such as the British Standards Institution (e.g., BS 8611:2016~\cite{bs86112016robots}), CIO Strategy Council (e.g., CAN/CIOSC 101:2019~\cite{ciosc}), IEEE Standard Association (e.g., IEEE 1232.3-2014~\cite{6922153}), and ISO/IEC Joint Technical Committee (e.g., ISO/IEC TR 24028:2020~\cite{iso_sa}). ISO/IEC JTC 1/SC 42~\cite{iso_sa1} committees are working on norms for AI including data (ISO/IEC JTC 1/SC 42/WG 2), fairness (ISO/IEC JTC 1/SC 42/WG 3), information security and data privacy (ISO/IEC JTC 1/SC 42/AHG 4). 

The IEEE Global Initiative on Ethics of Autonomous and Intelligent Systems also provides several standards including the model process for addressing ethical concerns during system design (IEEE 7000-2021~\cite{9536679}), transparency of autonomous systems (IEEE 7001-2021~\cite{winfield2021ieee}), and ontological standard for ethically driven robotics and automation systems (IEEE 7007-2021~\cite{9611206}). In 2021, NIST released a standard draft~\cite{stanton2021trust} for AI and user trust to help build trustworthy systems and understand user trust in AI, to minimize the risks and achieve benefits. The Association of Computer Machinery (ACM) Code of Ethics and Professional Conduct~\cite{acm_code} provides general principles, professional responsibilities, and professional leadership principles. These principles provide a basis/fundamental for ethical decision-making applied to a computing professional's conduct. Other national standards were defined by the Australian Computer Society (i.e., Code of Ethics and Professional Conduct) and the UK Committee on Standards in Public Life (i.e., Artificial Intelligence and Public Standards).

\subsubsection {AI Ethics Governance}

Ethics in corporate governance ensures trust and transparent in the exercise of power by making it credible not only to the shareholders, but also to stakeholders globally (e.g., employees, clients, suppliers)~\cite{dessain2008corporate}. Table~\ref{fig:aiethicsgovernance} presents 7 the ethical AI governance documents ranging from 2018 to 2021 and classified by name, author, organization, and year of publication. In 2019, the Berkman Klein Center for Internet and Society~\cite{ethics_govern_harvard} highlighted the following five key areas in which action is needed for ethics and governance of AI: 
the public dialogue about information quality of AI ethics, 
support of the development of inclusive AI governance frameworks, engagement of companies in developing socially beneficial paths for AI technology, and development of new approaches to teaching students and the public at large about the social implications of AI technology. 

The European Parliament~\cite{koene2019governance} also reviewed governance frameworks for algorithmic accountability and transparency such as principles vs. rules based governance, governance mechanisms (e.g., co-regulation by standards like GDPR, state intervention like funding research on transparent and accountable algorithms), and existing governance proposals such as algorithmic impact assessment-based solution
for accountability of algorithmic systems used by public authorities. They observed that most organisations rely on principles vs. rules that are more risks-oriented (i.e., minimization of harms/impact arising from the use of the AI technology), using impact assessment tools and they recommended a strong coordination between agencies when regulating algorithms for synchronization. A report on AI governance was also released by the University of Oxford~\cite{dafoe2018ai}. This report presents key points for an ideal AI governance: values and principles such as security and autonomy, and institutional mechanisms such as the ability to provide security, ensure safe AI, and ensure resilience to concept drift and hijacking.

In 2020, the school of governance of the Technical University of Munich (TUM)~\cite{tum_governance_2020} analyzed the relative weaknesses of major AI users and developers in terms of AI governance (e.g., heterogeneity in the multitude of principles and ethical AI solutions proposed in by organizations and governance) and identified three elements that are important for a proper embedding of ethics in AI governance: (1) hard and soft frameworks for regulating different AI use cases need to be distinguished by policy makers, (2) a global collaboration on defining minimum standards for AI must be done by stakeholder groups and policy makers, and (3) potential conflicts between different principles of AI ethics and the perceptions/preferences of stakeholders on them must be taken into account by corporate decision and policy makers. For ethical AI and autonomous system governance, the University of the West of England~\cite{8662743} states that all systems must be subjected to an ethical risk assessment following standards (e.g., BS 8611:2016) and then must be redesigned to reduce the impact of any ethical risks (e.g., redesign sub-systems for securely logging system inputs, outputs and decisions).

In 2019, AI4People~\cite{pagallo2019good} also proposed a Scalable, Modular, Adapted, Reflexive, and Technologically-savvy (SMART) coordination model using priority groups such as no-regret actions (i.e., new standards and procedures for AI), forms of engagement (e.g., new forum for discussion and consultation), and coordination mechanisms such as de-regulated special zones (i.e., kinds of living environments for the empirical testing and development of AI and robotics), in order to make rational decisions for critical issues.

\begin{table}[]
\centering
\caption{AI ethics algorithmic assessment tools}
\label{fig:aiethicsassess}
\begin{tabular}{|l|l|}
\hline
\textbf{Name} & \textbf{Organization} \\ \hline
AI Ethics Assessment Toolkit & \begin{tabular}[c]{@{}l@{}}Open Roboethics\\ Institute\end{tabular} \\ \hline
Algorithmic Impact Assessment & Canada \\ \hline
AI Impact Assessment & \begin{tabular}[c]{@{}l@{}}Platform for the \\ Information Society\end{tabular} \\ \hline
Ethics \& Algorithms Toolkit & San Francisco City \\ \hline
\begin{tabular}[c]{@{}l@{}}Algorithmic Impact Assessments: \\ A Practical Framework for Public \\ Agency Accountability\end{tabular} & AI Now \\ \hline
\begin{tabular}[c]{@{}l@{}}Algorithmic Accountability \\ Policy Toolkit\end{tabular} & AI Now \\ \hline
Algorithmic Equity Toolkit & ACLU of Washington \\ \hline
Ethical Self-assessment Tool & \begin{tabular}[c]{@{}l@{}}Centre for Applied\\ Data Ethics\end{tabular} \\ \hline
\begin{tabular}[c]{@{}l@{}}AI System Ethics \\ Self-Assessment Tool\end{tabular} & Smart Dubai \\ \hline
Ethical Assessment & \begin{tabular}[c]{@{}l@{}}Data for Children \\ Collaborative with UNICEF\end{tabular} \\ \hline
\begin{tabular}[c]{@{}l@{}}Algorithmic Impact Assessment\\ for the Public Interest\end{tabular} & Data Society \\ \hline
\begin{tabular}[c]{@{}l@{}}The Assessment List on\\ Trustworthy AI \end{tabular}& \begin{tabular}[c]{@{}l@{}}European Commission \\ HLEG AI\end{tabular} \\ \hline
Ethics Self-Assessment Tool & UK Statistics Authority \\ \hline
\begin{tabular}[c]{@{}l@{}}AI System Ethics \\Self-Assessment Tool\end{tabular} & \begin{tabular}[c]{@{}l@{}}University of Colorado \\Law School Legal Studies\\ Research\end{tabular} \\ \hline
\begin{tabular}[c]{@{}l@{}}Fair, Transparent, and Accountable\\ Algorithmic Decision-Making \\Processes\end{tabular} & L. Bruno et al. \\ \hline
Self-Reflection Tool & \begin{tabular}[c]{@{}l@{}}Responsible Research \\and Innovation \end{tabular}\\ \hline
\end{tabular}%
\end{table}

\begin{table*}[]
\centering
\caption{AI ethics governance}
\label{fig:aiethicsgovernance}
\begin{tabular}{|l|l|l|l|}
\hline
\textbf{Name} & \textbf{Author} & \textbf{Organization} & \textbf{Year} \\ \hline
ML Cards for D/MLOps Governance & I. Hellström & Databaseline & 2021 \\ \hline
AI Ethics and Governance. “Building a Connected, Intelligent and Ethical World” & C. Lütge & \begin{tabular}[c]{@{}l@{}}TUM School of \\ Governance\end{tabular} & 2020 \\ \hline
Ethics and Governance of AI &  & \begin{tabular}[c]{@{}l@{}}Berkman Klein Center\\ for Internet and Society\end{tabular} & 2019 \\ \hline
A governance framework for algorithmic accountability and transparency &  A. Koene et al. & European Parliament & 2019 \\ \hline
\begin{tabular}[c]{@{}l@{}}Machine Ethics: The Design and Governance of Ethical AI \\and Autonomous Systems\end{tabular} & A. F. Winfield et al. & \begin{tabular}[c]{@{}l@{}}University of the \\ West of England\end{tabular} & 2019 \\ \hline
\begin{tabular}[c]{@{}l@{}}On Good AI Governance 14 Priority Actions, a S.M.A.R.T. Model of Governance, \\ and a Regulatory Toolbox\end{tabular} & U. Pagallo & AI4People & 2019 \\ \hline
AI Governance: A Research Agenda & A. Dafoe & University of Oxford & 2018 \\ \hline
\end{tabular}%

\end{table*}

\begin{table*}[]
\centering
\caption{AI ethics checklists}
\label{fig:aiethicschecklist}
\begin{tabular}{|l|l|l|}
\hline
\textbf{Name} & \textbf{Organization} & \textbf{Year} \\ \hline
Deon & Driven Data & 2018-current \\ \hline
AI Fairness Checklist & Microsoft & 2020 \\ \hline
AlgorithmWatch & AlgorithmWatch & 2021 \\ \hline 
\begin{tabular}[c]{@{}l@{}}Checklist for Employers: Facilitating the Hiring of People with Disabilities Through\\ the Use of eRecruiting Screening Systems, Including AI\end{tabular} & \begin{tabular}[c]{@{}l@{}}EARN\end{tabular} & 2020 \\ \hline
\begin{tabular}[c]{@{}l@{}}Designing Ethical AI Experiences Checklist and Agreement\end{tabular} & \begin{tabular}[c]{@{}l@{}}CMU Software Engineering Institute\end{tabular} & 2019 \\ \hline
\begin{tabular}[c]{@{}l@{}}Trustworthy Repositories Audit \& Certification: Criteria and Checklist (TRAC)\end{tabular} & \begin{tabular}[c]{@{}l@{}}Center For Research \\ Libraries\end{tabular} & 2007 \\ \hline
\end{tabular}%
\end{table*}

\subsubsection{AI Ethics Assessment}
\label{assessment}

AI ethics assessment tools can be checklists or algorithmic processes that allows one to evaluate whether AI systems are ethically aligned. 

\paragraph{\textbf{Checklists}}

Table~\ref{fig:aiethicschecklist} shows 6 ethical AI checklists ranging from 2007 to now and classified by name, organization, and year of publication. Deon~\cite{deon_checklist} helps to easily build ethics checklist for data science projects. In 2019, the Carnegie Mellon University (CMU)'s Software Engineering Institute~\cite{cmu_checklist} proposed checklists with agreement topics 
(e.g., we will design our AI system with the following ethical principle in mind, we value respect and security) to guide the development of accountable and trustworthy AI systems. 

In 2020, Microsoft~\cite{microsoft_checklist} also proposed an AI fairness checklist to address fairness issues in different stages of the AI project lifecycle: planning meetings (e.g., envision system and scrutinize system vision for potential fairness-related harms), reviews (e.g., define and scrutinize system architecture considering potential fairness-related harms), prototyping (e.g., fairness evaluation in system/dataset prototype), building (e.g., fairness evaluation in production datasets/code/system), launching (e.g., revision of systems to mitigate any harms), and evolution (e.g., continuous updates of checklists, continuous monitoring of fairness in systems to identify and correct new biaises).

In 2021, AlgorithmWatch~\cite{algowatch-report} also defined two checklists (i.e., transparency triage and transparency report) to help identify which ethical transparency issues must be addressed/documented before, during, and after the usage of an automated decision-making system w.r.t ethic principles such as justice, fairness, privacy, cybersecurity, autonomy, and transparency.

In 2007, the Center For Research Libraries~\cite{crl_checklist} proposed checklists for the audit and certification of organizational infrastructures, digital object management, technologies, technical infrastructure, and security based on evidence, transparency, adequacy, and measurability criteria. This checklists help to address the trustworthiness of data management systems (including digital repositories), which is important to ensure the safety and security of data processed by AI systems. The Employer Assistance and Resource Network (EARN)~\cite{earn_disa} on disability inclusion also defined a set of questions for leadership (e.g., compliance with diversity and inclusion policies), human resources personnel (e.g., inclusion of individuals with disabilities), equal employment opportunity managers (e.g., equity and diversity inclusion), and procurement officers (e.g., nondiscrimination/fairness for people with disabilities). These questions on diversity and inclusion are important for AI governance in order to foster ethical AI systems.

\paragraph{\textbf{Algorithmic assessment}}

Table~\ref{fig:aiethicsassess} shows 16 ethical AI algorithm assessments found in the literature, classified by name and description. The Open Roboethics Institute~\cite{faie-report} proposed an algorithmic assessment tools consisting in 3 phases: identification of use cases and stakeholders (e.g., use of machine learning algorithm for better prediction of nanny profiles, professional nanny as stakeholder), ethic risk discovery (e.g., using societal values such as transparency, trust, fairness, diversity, human rights), and creation of an implementation road map w.r.t the societal values. The Canadian government also built an algorithmic impact assessment tool~\cite{ethics_canada} to mitigate the impact associated to the use of an automated AI decision system. 

The ECP Platform for the Information Society~\cite{ecp-report} developed steps to conduct an AI impact assessment (e.g., determine the need to perform an AI assessment, describe the AI application and its benefits, checking whether application is reliable/safe/transparent). San Francisco City provides an algorithm toolkit\footnote{https://ethicstoolkit.ai/} consisting of two parts: an algorithm risk assessment (e.g., impact, data use risk, accountability risk, third-party methodology risk, historic bias risk, technical bias risk) and an algorithm risk management that matches risk to mitigation, using various severity levels such as very low, low, moderate, high, and extreme. 

The AI Now Institute~\cite{ainow_rep} also released an algorithmic impact assessment process consisting of 5 steps: pre-acquisition (e.g., procurement planning, contract negotiation and signing), disclosure requirements (e.g., automated decision system disclosure, potential harms/bias/inaccuracy evaluation), comment period (e.g., public engagement to the process), due process challenge period (e.g., the public challenges the process to identify weaknesses), and impact assessment renewal. Other algorithmic assessment tools were defined by the American Civil Liberties Union (ACLU) of Washington~\cite{aclu_2020}, Smart Dubai~\cite{ethicsai_tool}, the Data for Children Collaborative with UNICEF~\cite{unicef_report}, the Data Society~\cite{moss2021assembling}, the European Commission High-Level Expert Group (HLEG) AI~\cite{hleg_ai}, the UK Statistics Authority~\cite{uk-report}, the University of Colorado Law School Legal Studies Research~\cite{kaminski2020algorithmic}, and the European Commission Responsible Research and Innovation~\cite{lepri2018fair}.

\subsubsection{AI Ethics Software}
\label{pract-tools}


Table~\ref{fig:aiethicstools} shows 28 ethical AI softwares found in the literature, classified by name and description. Aequitas~\cite{aequitas_checklist} is a fairness audit software that allows one to audit machine learning (ML) models for discrimination and bias to make equitable decisions. 
Deon~\cite{deon_checklist} is another tool that allows one to build ethical checklists that an AI application must follow. Fairness, Accountability, and Transparency (FAT) Forensics~\cite{fat_tool} is a toolbox for the evaluation of FAT principles in predictive systems. AI Fairness 360\footnote{https://aif360.mybluemix.net/} is tool of metrics developed by IBM to check, report, and mitigate discrimination and bias in ML models. IBM also developed AI Explainability 360\footnote{https://aix360.mybluemix.net/} to help explain/interpret ML models and AI FactSheets 360\footnote{https://aifs360.mybluemix.net/} to improve transparency by collecting relevant information (facts) such as the criticality of the model, metrics about dataset, model, service or actions taken during the deployment of the model/service. 

Other tools such as Fairlearn~\cite{fairlearn_tool}, What-If~\cite{whatif_tool}, TensorFlow (TF) Fairness~\cite{fairness_ind}, LinkedIn Fairness~\cite{lif_tool},  ML-fairness-gym~\cite{ml_fairness_gym}, and UnBias Fairness~\cite{unbias_tool} allow one to check and improve the fairness of ML models. Audit AI~\cite{auditai_tool}, TF Model Analyzer/Remediation~\cite{tfrem_tool}, and REVISE~\cite{revise_tool} tools also help to build unbiased ML solutions. Tools like SHAP~\cite{shap_tool}, LIME~\cite{lime_tool}, InterpretML~\cite{interpret_tool}, and PAIR Facets~\cite{pair_tool} help to explain and interpret data and ML models. Tools such as Adversarial Robustness Toolbox (ART)~\cite{art_tool}, PySyft~\cite{pysyft_tool}, and TF Privacy~\cite{tfpriv_tool} help secure models and data during ML lifecycle. Data Version Control (DVC)~\cite{dvc_tool} allows one to track and control the versioning of data and models (similar to Git for source code management).

\begin{table}[]
\centering
\caption{AI ethics softwares}
\label{fig:aiethicstools}
\begin{tabular}{|l|l|}
\hline
\textbf{Name} & \textbf{Description} \\ \hline
Aequitas & Bias \& Fairness Audit Toolkit \\ \hline
Deon & Checklist for data science projects \\ \hline
eXplainability Toolbox & Algorithmic bias and explainability \\ \hline
FAT Forensics & \begin{tabular}[c]{@{}l@{}}evaluating Fairness, Accountability \\and Transparency of AI systems\end{tabular} \\ \hline
AI Fairness 360 & \begin{tabular}[c]{@{}l@{}}An Open Source Toolkit for detection \\ and mitigation of bias in ML models\end{tabular} \\ \hline
AI Explainability 360 & \begin{tabular}[c]{@{}l@{}}An Open Source Toolkit for \\interpretability and explainability of\\ datasets and ML models\end{tabular} \\ \hline
AI FactSheets 360 & \begin{tabular}[c]{@{}l@{}}Trust in AI by increasing trans- \\parency and enabling governance\end{tabular} \\ \hline
Fairlearn & \begin{tabular}[c]{@{}l@{}}Toolkit for assessing and improving \\fairness in AI\end{tabular} \\ \hline
Algofairness & BlackBox Auditing Tool \\ \hline
FairSight & Fair Decision Making Tool \\ \hline
PAIR Facets & \begin{tabular}[c]{@{}l@{}}Analyzing ML datasets \\ by visualization\end{tabular} \\ \hline
What-if Tool & Playing with AI Fairness \\ \hline
ML-fairness-gym & \begin{tabular}[c]{@{}l@{}}A Tool for Exploring Long-Term \\Impacts of ML Systems\end{tabular} \\ \hline
InterpretML & Explain blackbox ML \\ \hline
LinkedIn Fairness Toolkit & \begin{tabular}[c]{@{}l@{}}Measurement of fairness in large \\ scale ML workflows\end{tabular} \\ \hline
Audit AI & \begin{tabular}[c]{@{}l@{}}Detect demographic differences in \\ the output of ML models\end{tabular} \\ \hline
SHAP & SHapley Additive exPlanations \\ \hline
LIME & \begin{tabular}[c]{@{}l@{}}Local Interpretable Model-agnostic \\ Explanations\end{tabular} \\ \hline
TF Fairness Indicators & \begin{tabular}[c]{@{}l@{}}Easy computation of  fairness metrics \\for binary and multi-class classifiers\end{tabular} \\ \hline
TF Model Analyzer & Performing model evaluation \\ \hline
TF Model Remediation & \begin{tabular}[c]{@{}l@{}}Reduces or eliminates user \\harm resulting from underlying\\ performance biases\end{tabular} \\ \hline
Skater & Model Interpretation \\ \hline
UnBias Fairness Toolkit & Critical and civic thinking tool \\ \hline
REVISE & \begin{tabular}[c]{@{}l@{}}A Tool for Measuring and Mitigating \\ Bias in Visual Datasets\end{tabular} \\ \hline
ART & \begin{tabular}[c]{@{}l@{}} Adversarial Robustness Toolbox\end{tabular} \\ \hline
PySyft &  \begin{tabular}[c]{@{}l@{}} ML Privacy using Federated \\Learning,Differential Privacy, and \\Encrypted Computation\end{tabular} \\ \hline
TF Privacy &  \begin{tabular}[c]{@{}l@{}} Training ML models with \\Privacy for training data\end{tabular} \\ \hline
DVC & \begin{tabular}[c]{@{}l@{}}Versioning and control of \\data and models\end{tabular}\\ \hline
\end{tabular}%
\end{table}

\subsubsection{AI Ethics Awareness}
\label{awareness}

Table~\ref{fig:aiethicstraining} shows 10 ethical AI awareness training found in the literature, classified by name and organization. Several programs are led by universities and companies to train practitioners, researchers, and citizens on ethical AI principles, guidelines, and applications. The University of Montreal offers a Massive Open Online Course (MOOC) on \textit{Bias and Discrimination in AI}~\cite{udem_eai}: types of bias, fairness metrics, public laws and policies, technical/institutional bias and discrimination mitigation. The University of Helsinki also provides a course on Ethics of AI; it defines ethics and its principles such as non-maleficence, accountability, transparency, human rights, and fairness. 

\begin{table}[]
\caption{AI ethics - Online training}
\label{fig:aiethicstraining}
\begin{tabular}{|l|l|}
\hline
\textbf{Name} & \textbf{Organization} \\ \hline
Data Science Ethics & University of Michigan \\ \hline
Secure \& Private AI & Udacity \\ \hline
Ethics in AI and Data Science & Linux Foundation \\ \hline
Ethics of AI & University of Helsinki \\ \hline
AI Ethics for Business & Seattle University \\ \hline
Bias and Discrimination in AI & Université de Montréal \\ \hline
Practical Data Ethics & Fast AI \\ \hline
\begin{tabular}[c]{@{}l@{}}Data Ethics, AI and \\Responsible Innovation\end{tabular} & University of Edinburgh \\ \hline
\begin{tabular}[c]{@{}l@{}}Ethics for Engineers: Artificial \\Intelligence\end{tabular}& MIT \\ \hline
\begin{tabular}[c]{@{}l@{}}Explainable AI: Scene Classification \\and GradCam Visualization\end{tabular} & Coursera \\ \hline
\end{tabular}%
\end{table}

The Seattle University course on \textit{AI ethics for business}~\cite{seattle_eai} instructs students on ethical issues around the AI technologies and then provides possible mitigations to them. The course \textit{Data Ethics, AI and Responsible Innovation}~\cite{ueding_eai} is also given by the University of Edinburgh. It helps to understand societal/political/legal/ethical issues around data lifecycle and ethical concepts (e.g., responsibility, data governance) and it also helps to identify, assess, evaluate ethics issues in data science and industry. Massachusetts Institute of Technology (MIT) also provides a course titled \textit{Ethics for Engineers: Artificial Intelligence}~\cite{mit_eai}. The course help engineers to understand ethics in different cases such as AI, Social media and data, Engineering and society broadly, Engineering and politics broadly, Engineering and democracy. Other courses are provided by training companies such as Coursera, Udacity, and Fast AI to help implement ethical AI into AI projects (e.g., secure and private AI, explainable AI). 

\subsection{What are the gaps between ethical AI principles and their implementations as well as their root causes?}

After identifying current AI ethics implementations, we found several gaps between ethical AI principles and their implementations~\cite{tum_governance_2020}. In this section, we report about 5 
gaps identified between ethical AI principles and their implementations, as well as their root causes (RQ3). 
\subsubsection{Principle-Implementation Gaps} From the analysis of the literature, we have identified the following gaps:

\paragraph{\textbf{Lack of implementation tools for some ethical AI principles}} Several open source tools for AI ethics have been developed including those in Table~\ref{fig:aiethicstools}. However, most tools were focused on few principles such as Fairness, Non-discrimination, and Bias (see Table~\ref{fig:aiethicstools}). Few tools targeted Explainability, Interpretability, Security, and Privacy principles (e.g., AI Explainability 360, LIME,  ART, TF Privacy). In addition, the existing tools do not even cover all aspects of the principles targeted; for example, the fairness implementation is limited to input data, AI models, and outputs without taking into account elements such as the fairness of the AI design/engineering process, the fairness of the AI libraries/tools and infrastructure used in the process, and EthicsOps (i.e., the continuous integration and monitoring of ethics principles in the AI project lifecycle for harm prevention). On the 33 principles, there are missing tools for several principles including Accountability, Responsibility, Traceability, Controllability, Equity, Justice, Constestability, Reliability, Autonomy, Sustainability, Governance, Integrity, Trustworthy, Transparency (partially covered by explainability/interpretability but no traceability), and Inclusiveness. 

\begin{boxblock}{Summary 4}
    \begin{itemize}
      \item The existing ethical guidelines promote education and awareness to foster an ethical mind-set, the use of fairness and non-discrimination in products and models as well as data quality control, diversity and inclusiveness of team roles, explainability of model functions and outputs, traceability of all training and operating data.
      \item The current codes of conduct include laws such as Canada (e.g., Privacy Act), United States (e.g., Privacy Protection Act), Europe (e.g., GDPR), Australia (e.g., Privacy Act), China (e.g., Personal Information Protection Law), and South Africa (e.g., Protection of Personal Information Act); and standards such as Bristish Standards Institution (e.g., BS 8611:2016), CIO Strategy Council (e.g., CAN/CIOSC 101:2019), IEEE (e.g., IEEE 7001-2021), ISO/IEC (e.g., ISO/IEC TR 24028:2020), and NIST (e.g., NISTIR 8332).
      \item The existing ethical governance frameworks promote the development of inclusive ethical AI governance, strong collaboration and coordination between stakeholder groups and policy makers, the use of algorithmic impact assessment-based solutions for accountability of algorithmic systems.
      \item The current ethical assessments include checklists such as Deon, Microsoft AI Fairness, Center for Research Libraries~\cite{crl_checklist}, and EARN~\cite{earn_disa}; and algorithmic assessments such as Open Roboethics Institute~\cite{faie-report}, Data Society~\cite{moss2021assembling}, UK Statistics Authority~\cite{uk-report}, European Commission HLEG AI~\cite{hleg_ai}, and UNICEF~\cite{unicef_report}.
      \item The existing ethics tools include Aequitas, Deon, SHAP, LIME, AI FactSheets 360, AI Fairness 360, AI Explainability 360, PAIR Facets, What-If Tool, InterpretML, PySyft, Adversarial Robustness Toolbox, Data Version Control, Tensorflow Privacy.
      \item The list of institutions currently providing ethical education and awareness materials include : Universite de Montreal (course: \textit{Bias and Discrimination in AI}), Seattle University (course: \textit{AI ethics for business}), University of Edinburgh (course: \textit{Data Ethics, AI and Responsible Innovation}), and MIT (course: \textit{Ethics for Engineers: Artificial Intelligence})
    \end{itemize}
\end{boxblock}

\paragraph{\textbf{Lack of effective standards}} There are few ethical AI standards~\cite{oceanis} that cover ethical principles and some are still at the stage of draft~\cite{o2019legal, mittelstadt2019ai}. In the studied ethical AI implementations, most standards address AI ethics in terms of governance policies with limited work on technical policies, i.e., step-by-step instructions to apply ethical principles (e.g., transparency, traceability, accountability) during the AI project's lifecycle (i.e., planning, implementation, delivery, and monitoring). 

\paragraph{\textbf{Lack of practical training courses on AI ethics}} 
Universities and training companies proposed generic training courses on \textit{what} are AI ethics, guidelines, standards, and very few on \textit{how} AI ethics is (and should be) really applied during the design and development process of an AI system. There is a need for consistent and up-to-date courses on the application of AI ethics in practice.

\paragraph{\textbf{Weakness of 
the implementation of AI ethics principles in corporate governance}} Corporate governance of AI have issues such as improper supervision and monitoring, deficient auditing of financial reports, decision making process issues (e.g., conflicts, divergence), and flaws in multilateral contracts. After analyzing the existing ethical AI governance, very few documents addressed the application of ethical principles to fix these issues, precisely in terms of transparency (e.g., in the auditing of documentation and financial reports, effective application of ethics guidelines and laws), responsibility (e.g., social duty, organisational), equity (e.g., in the decision-making processes with both shareholders and stakeholders, employees and managers), autonomy (e.g., free will of stakeholders), and sustainability (e.g., continuous adaptation of products/services to the market).

\paragraph{\textbf{lack of coverage of ethics
in artificial general intelligence by implementation
materials}} Like T. Hagendorff  \cite{hagendorff2020ethics}, we found that ethics in Artificial General Intelligence (AGI) and its consequences on humanity are not discussed in implementations materials, while most ethics AI principles cover broad aspects of AGI. It is understood that we are currently at the stage of artificial narrow intelligence (weak AI) and have not yet reached artificial super intelligence~\cite{yampolskiy2015artificial}, but AI is maturing fast and there is a need to anticipate potential ethical issues and consequences~\cite{bostrom2002existential} that AI may pose in the near future (e.g., controllability, transhumanity).


\subsubsection{Root causes of principle-implementation gaps} After analysing the implementations and gaps, the following causes of gaps have been identified.

\paragraph{\textbf{Human bias}} 

The studied ethical AI implementations does not provide clear translation from the abstract ethical principles (shared by humans and the society) to ML algorithms and tools. The translation is distorted 
because (1) most people involved in the implementations are technical (i.e., with a computer science background) and their individual perception~\cite{arvan2018mental} limits the implementation, (2) there is lacking proethically designed human-computer interaction or networks of AI systems~\cite{morley2021initial}, and (3) it is hard to translate complex human (ethical) values into design tools that are easy to use; which makes them hardly actionable~\cite{vakkuri2019ethically}.

\paragraph{\textbf{Poor diversity within the AI community}} The diversity problem is still present in AI research and policy-making teams: (1) all races are not represented in those deciding AI groups (i.e., mainly consisting of white men), (2) there is no equity about genders in the deciding groups (i.e., mainly men and only few women), (3) no inclusion of people with disabilities, and (4) no skill balance (i.e., mainly technical people with science backgrounds). This lack of diversity makes ethical AI implementations (e.g., guidelines, laws, standards, tools) unaligned with ethical AI principles. 

\paragraph{\textbf{Complexity, interconnection of decisions, and processes that are learned from data}} These implementation losses~\cite{koene2019governance} were addressed by the European Parliament. Complexity expresses the behavior of common ML algorithms. For example, ML algorithm modules execute different tasks and the combination provides final results. However, each task has different meanings for data and the randomness of ML algorithms make a step-by-step explanation more challenging~\cite{koene2019governance}. In addition, decisions are often the result of multiple optimization (a batch of simultaneous solutions), that make explainability and interpretability difficult. Furthermore, the ML models that learnt form data are not understandable and transparent. In the context of online learning, the models are continuously updated with data; thus, the application of ethical AI principles must be continuous, or monitored to be effective. In addition, assigning responsibility to ML algorithm harms is difficult and this can lead to issues with moral responsibility~\cite{vakkuri2019ethically, morley2021initial}.

\begin{boxblock}{Summary 5}
      \begin{itemize}
      \item The 5 identified gaps are : \textit{the lack of implementation tools for some ethical principles} (i.e., many ethics principles including Accountability/Traceability/Controllability are not supported, no 
      EthicsOps), \textit{the lack of effective standards} (i.e., limited work on the step-by-step application of ethics principles in the AI lifecycle), \textit{the lack of practical training} (i.e., lack of practical courses on AI ethics), \textit{the weakness of Ethics in corporate governance} (e.g., lack of transparency in the documentation and auditing of financial reports, lack of equity in decision making processes, lack of sustainability for services and products), and \textit{lack of coverage of ethics in artificial general intelligence by implementation materials} (i.e., no discussion about this topic and its consequences~\cite{bostrom2002existential}).
      \item The 5 identified root causes are \textit{human} (i.e., individual perception bias~\cite{arvan2018mental}, team selection bias, no proethically designed human-computer interaction~\cite{morley2021initial}, translation complexity of ethical values~\cite{vakkuri2019ethically}), \textit{diversity} (i.e., no balance about races/genders, no disability inclusion), \textit{complexity, interconnection of decisions, and processes learned from data} (i.e., step-by-step explanation is hard due to the randomness of ML algorithms, multiple optimizations, and hiddenness of ML models), \textit{public-private partnerships and industry-funded research} (i.e., AI democratization issue, AI research freedom~\cite{hagendorff2020ethics}), and \textit{lack of metrics} for assessing the implementation of AI ethics principles~\cite{thomas2020reliance, eitel2021beyond}.
      \end{itemize}
\end{boxblock}

\paragraph{\textbf{Issue in public–private partnerships and industry-funded research, in the field of AI}} In the studied ethical AI implementations, there are several actors such as inter-governmental partnerships (e.g., OECD, G20), international partnerships (e.g., GPAI, IEEE), national (e.g., Montreal Declaration for a Responsible AI), and universities (e.g., Universite de Montreal, Eindhoven University, ETH Zurich). Most universities are funded by public-private partnerships that limits their freedom in AI research~\cite{hagendorff2020ethics} and the democratization of AI; since some results of the research are often proprietary and the research is more focused on the needs of funding partners than on solving real problems in AI. Consequently, these issues affect the implementation of models and support tools for AI ethics. 

\paragraph{\textbf{Lack of metrics for ethical AI principles}}
There is missing proper models and metrics to evaluate the application of those principles~\cite{thomas2020reliance, eitel2021beyond}. Few metrics have been proposed including fairness~\cite{corbett2018measure}, privacy~\cite{mohassel2017secureml}, and robustness~\cite{nicolae2018adversarial} but they are not enough and does not capture well the abstraction of ethical principles. In addition, some work such as~\cite{sablayrolles2020radioactive, bdcc5020020} tried to provide models for ML traceability but they are not yet effective.





\section{Gap Mitigation}~\label{mitigation}


In this section, we provide recommendations to mitigate the identified gaps (RQ4). 

\textbf{Recommandation 1}. Several actors (from different genders, races, disabilities, skills) such as policy markers, customers, philosophists, compliance specialists, security experts, software architects and developers must be involved in the design and implementation of the ethics tools, not only technical people (i.e., security experts, software architects, developers), to ensure an effective implementation and application of ethical principles. Some researches have been done to define various metrics for fairness~\cite{corbett2018measure,10.1145/3457607}, discrimination and bias~\cite{10.1145/3457607}, privacy~\cite{abadi2016deep, sweeney2002k, mohassel2017secureml, bonawitz2017practical, 7950921, 10.1145/775047.775089}, security~\cite{bonawitz2017practical,barreno2010security,papernot2016towards,mohassel2017secureml}, robustness~\cite{7958570,nicolae2018adversarial,brendel2017decision,papernot2016distillation,moosavi2017universal,papernot2016technical,brendel2017decision,rauber2017foolbox}, explainability~\cite{ribeiro2016should,lundberg2017unified,carvalho2019machine}, interpretability~\cite{ribeiro2016should,lundberg2017unified,carvalho2019machine}, transparency and accountability~\cite{mitchell2019model,ribeiro2016should,lundberg2017unified,carvalho2019machine}, traceability and auditability~\cite{weng2019deepchain,chen2018traceability,mora2021traceability}. However, more metrics need to be developed to evaluate whether AI systems are ethically-aligned. 

\textbf{Recommandation 2}. Teams must aligned projects with EthicsOps to ensure that solutions built follow ethics principles. EthicsOps ensures the continuous application and control of ethics principles during the project lifecycle (planning, execution, delivery). During the planning phase, business requirement specifications must follow existing ethics laws, standards, and guidelines (see Section~\ref{existing-guidelines}). The internal ethics policies of the organization 
must be updated continuously, since ethical standards and guidelines evolve, as more research results are produced 
(e.g., models, techniques) in the AI ethics field. In addition, team selection and change management must take into account diversity, inclusiveness, non-discrimination, and equity principles (i.e., equity irrespective of the gender, race, and disability). 

Ethical controls using checklists, Algorithmic Impact Assessment (AIA), Key Ethics Indicators (KEI) based on ethics metrics~\cite{corbett2018measure,10.1145/3457607,abadi2016deep, sweeney2002k, bonawitz2017practical,barreno2010security,7958570,nicolae2018adversarial,ribeiro2016should,lundberg2017unified,weng2019deepchain} must be also enforced during the planning, implementation, and delivery of AI projects. During the execution and delivery phases, ethics by design practices~\cite{eu-ethos} help to proactively (by using the principles as requirements) prevent ethical issues from arising. For example, Transparency by design and Fairness by design ensure that outcomes of the project and the project itself are transparent, fair, and auditable by independent third parties (e.g., XAI from EthicalML). In an IT project, EthicsOps can be coupled with existing practices (e.g., DevSecOps~\cite{myrbakken2017devsecops}, AIOps~\cite{dang2019aiops}) to design, build, deploy, and monitor ethically-aligned AI systems.

\textbf{Recommandation 3}. More partnerships between public-private sectors and universities must be created to accelerate research, development, and knowledge transfer in AI ethics. For example, Facebook (now Meta) partners with the Technical University of Munich (TUM) to support the creation of an independent AI ethics research center with an investment of 7.5 millions USD~\cite{meta_partnership}. In Canada, the Quebec and federal governments invested together an amount of 15 millions CAD in total for the creation of the International Centre of Expertise in Montreal on Artificial Intelligence (CEIMIA)\footnote{https://ceimia.org/en/} to foster research and innovation on responsible AI based on ethical principles, human rights, inclusion, diversity, innovation, and economic growth. In addition, research agreements between universities and companies must be clear about time investment (e.g., 50/50), confidentiality, ownership, commercialization rights, retained rights, compliance with laws, publications, warranties, indemnification and liability, and termination. 

\begin{boxblock}{Summary 6}
    To mitigate the gap between principles and implementations, the solutions include: 
    \begin{itemize}
      \item Inclusiveness and diversity of team roles (races, genders, disabilities, skills).
      \item Education and awareness on ethical values, culture, methods, and practices~\cite{taebi2019importance} to foster an ethical mindset (see Summary 4).
      \item Increase of partnerships between private-public sectors and universities to accelerate research on AI ethics. 
      \item Application of existing ethics laws and standards during the AI governance and engineering process while staying up-to-date with new changes (see Summary 4).
      \item Application of EthicsOps in combination with existing practices (e.g., DevSecOps~\cite{myrbakken2017devsecops}, AIOps~\cite{dang2019aiops}) to design, build, deploy, and monitor ethically-aligned AI systems from staging to the production environment.
    \end{itemize}
\end{boxblock}

\textbf{Recommandation 4}. More programs on human-centered research, and AI ethics must be included in universities and engineering schools, to instill and foster the adoption of ethical values, culture, methods, and practices~\cite{taebi2019importance} by future scientists and engineers. For example, universities such as the University of Montreal, Massachusetts Institute of Technology, Seattle University, and University of Michigan proposed some courses on AI ethics (see Table~\ref{fig:aiethicstraining}). Scientists and engineers need to know AI ethics in practice and work in an inclusive/diverse environment to build effective and sustainable tools to prevent AI systems from harming. Existing Standards (presented in Summary 4) must be followed during the AI engineering process and these standards must be continuously updated. National and international organizations need to stay abreast of new changes (i.e., new regulatory norms, standard updates).

\section{Threats to validity}\label{threat2valid}

In this work, we have analyzed 350 reports and websites, and retained 100 reports containing AI principles and AI implementations. The 250 remaining that did not contained AI principles or implementations were ignored. This process was done manually and we may have missed information containing AI principles or implementations. 

In addition, the extraction of principles and implementations was done manually 
with basic functions of Microsoft Excel. It took 2 months to search, read, 
identify, extract, and record relevant information from the 100 reports and websites. Nevertheless, we have verified document sources and the online publishers to ensure that they are trustworthy. During the extraction process, we have repeated statistics 2 times to avoid any miscalculation. 

\section{Conclusion}~\label{conclusion}

In this work, we have analyzed ethical AI principles and their implementations as well as the principle-implementation gaps. We have also proposed some recommendations to mitigate the gaps. Results show that Transparency is the most cited principle among the studied countries. The 11 global principles that cover most continents with a high number of occurrences are Privacy, Transparency, Fairness, Security, Safety, Responsibility, Accountability, Explainability, Well-being, Human Rights, and Inclusiveness. The ethical AI principles published by GPAI, OECD, and G20 are missing the Responsibility principle, compared to the global principles. United States, United Kingdom, and Canada released the highest number of principles between 2016 and 2021. After reviewing more than 121 ethical implementations, 5 principle-implementation gaps were identified including lack of effective implementation tools, lack of effective standards, and weakness of the implementation of AI ethics principles in corporate governance. In addition, 5 root causes of gaps were also found including human bias, lack of metrics for assessing the implementation of ethical principles, poor diversity and inclusion. Recommendations for gap mitigation include: inclusiveness and diversity of team roles, partnerships between public-private sectors and universities for fostering AI ethics research, and the use of EthicsOps. In the future, we plan to develop some ethical tools and propose practical assessment mechanisms, to bridge the gap between ethical AI principles and their implementations.

\section*{Acknowledgment}

This work is partly funded by the Fonds de Recherche du Québec (FRQ), Natural Sciences and Engineering Research Council of Canada (NSERC), Canadian Institute for Advanced Research (CIFAR), and Mathematics of Information Technology and Complex Systems (MITACS).

\bibliographystyle{IEEEtran}
\bibliography{bibliography}
\end{document}